\documentclass[aps,prd,showpacs,superscriptaddress,preprintnumbers,dblfloatfix,nofootinbib,floatfix,twocolumn,12points]{revtex4-1}
\usepackage{graphicx} 
\usepackage{amsfonts,amsmath,amssymb}
\usepackage{tikz}
\usepackage{makecell}
\usepackage{float}
\usepackage{environ}
\usetikzlibrary{decorations.pathmorphing,decorations.markings,trees,shapes}
\usepackage[force]{feynmp-auto}
\tikzset{
    photon/.style={decorate, decoration={snake}},
    electron/.style={draw=black, postaction={decorate},decoration={markings,mark=at position .55 with {\arrow[draw=black]{>}}}},
        antielectron/.style={draw=black, postaction={decorate},decoration={markings,mark=at position .55 with {\arrow[draw=black]{<}}}},
    gluon/.style={decorate, draw=magenta,
        decoration={coil,amplitude=4pt, segment length=5pt}} 
}

\usepackage{wasysym}
\usepackage{bm}
\usepackage{hyperref}
\usepackage{latexsym}
\usepackage{textcomp}
\usepackage{subcaption}

\usepackage{verbatim}
\usepackage{slashed}
\usepackage{array}
\usepackage{multirow} 
\usepackage{color}
\usepackage{dsfont}
\usepackage{tikz-feynman}
\usepackage{feynmp-auto}

\definecolor{myred}{rgb}{0.7, 0, 0}
\definecolor{myblue}{rgb}{0, 0, 0.7}
\definecolor{mygreen}{rgb}{0.04, 0.7, 0.5}
\definecolor{mygray}{rgb}{0.1, 0.1, 0.1}

\def\bea  {\begin{eqnarray}}   \def\eea  {\end{eqnarray}}
\def\bean {\begin{eqnarray*}}  \def\eean {\end{eqnarray*}}
\def\nn{\nonumber}
\def\bry{\begin{array}}
\def\ery{\end{array}}
\def\TeV{\,{\rm TeV}}
\def\GeV{\,{\rm GeV}}
\def\MeV{\,{\rm MeV}}
\def\keV{\,{\rm keV}}
\def\eV{\,{\rm eV}}
\def\Mpl{ M_{\rm Pl}}

\def\eps{\epsilon}
\def\Lqcd{\Lambda_{\rm QCD, k\ne 0}}

\def\ephi{\epsilon_\phi}

\def\mapi{m^\pi_a}
\def\aoscpi{a_{{\rm osc},\pi}}

\newcommand{\ie}{{\it i.e.}}
\newcommand{\eg}{{\it e.g.}}

\newcommand{\skipnew}[1]{}

\usetikzlibrary{decorations.pathmorphing}
\usetikzlibrary{shapes.geometric}
\newcommand{\beq}{\begin{equation}}
\newcommand{\eeq}{\end{equation}}
\newcommand{\gsim}{\lower.7ex\hbox{$\;\stackrel{\textstyle>}{\sim}\;$}}
\newcommand{\lsim}{\lower.7ex\hbox{$\;\stackrel{\textstyle<}{\sim}\;$}}

\newcommand{\cO}{\mathcal{O}}
\newcommand{\cL}{\mathcal{L}}

\newcommand{\fac}{(f_a)_{\rm crit}}

\numberwithin{equation}{section}
\renewcommand{\theequation}{\arabic{equation}}

\def\UMD{\small{Maryland Center for Fundamental Physics, University of Maryland, College Park, MD 20742, USA}}

\usepackage{epsfig}
\usepackage{placeins}
\usepackage{epstopdf}
\usepackage{braket}
\usepackage[export]{adjustbox}

\usepackage{subcaption}

\makeatletter
\renewcommand\@makecaption[2]{%
  \par
  \vskip\abovecaptionskip
  \begingroup
    \small\rmfamily
    \begingroup
      \samepage
      \flushing
      \let\footnote\@footnotemark@gobble
      \@make@capt@title{#1}{#2}\par
    \endgroup
  \endgroup
  \vskip\belowcaptionskip
}
\makeatother

\def\mZ{\mathbb{Z}}
\def\epsphi{\epsilon_{\phi}}

\definecolor{myred}{rgb}{0.7, 0, 0}
\definecolor{myblue}{rgb}{0, 0, 0.7}
\definecolor{mygreen}{rgb}{0.04, 0.7, 0.5}
\definecolor{mygray}{rgb}{0.1, 0.1, 0.1}

\hypersetup{colorlinks,citecolor=myred,linkcolor=myblue,urlcolor=myblue,linktocpage=true}

\def\be   {\begin{equation}}   \def\ee   {\end{equation}}
\def\ba   {\begin{array}}      \def\ea   {\end{array}}
\def\bea  {\begin{eqnarray}}   \def\eea  {\end{eqnarray}}
\def\bean {\begin{eqnarray*}}  \def\eean {\end{eqnarray*}}
\def\nn{\nonumber}
\def\bry{\begin{array}}
\def\ery{\end{array}}
\def\TeV{\,{\rm TeV}}
\def\GeV{\,{\rm GeV}}
\def\MeV{\,{\rm MeV}}
\def\keV{\,{\rm keV}}
\def\eV{\,{\rm eV}}
\def\Mpl{ M_{\rm Pl}}

\def\Hrh{H_{\rm rh}}

\def\Trh{T_{\rm rh}}

\def\trh{t_{\rm rh}}

\def\cL{\mathcal{L}}
\def\Lnp{\Lambda_{\rm NP}}
\def\mqcd{(m_a)_{\rm QCD}}
\def\mqcds{(m^2_a)_{\rm QCD}}
\newcommand{\vev}[1]{\left\langle #1 \right\rangle}

\newcommand{\bl}{\left}
\newcommand{\br}{\right}

\numberwithin{equation}{section}
\renewcommand{\theequation}{\arabic{section}.\arabic{equation}}

\def\UMD{\small{Maryland Center for Fundamental Physics, University of Maryland, College Park, MD 20742, USA}}

\usepackage{tikz,xcolor,hyperref}
\definecolor{lime}{HTML}{A6CE39}
\DeclareRobustCommand{\orcidicon}{%
	\begin{tikzpicture}
	\draw[lime, fill=lime] (0,0) 
	circle [radius=0.16] 
	node[white] {{\fontfamily{qag}\selectfont \tiny ID}};	\draw[white, fill=white] (-0.0625,0.095) 
	circle [radius=0.007];	\end{tikzpicture}
	\hspace{-2mm}}
\foreach \x in {A,...,Z}{%
	\expandafter\xdef\csname orcid\x\endcsname{\noexpand\href{https://orcid.org/\csname orcidauthor\x\endcsname}{\noexpand\orcidicon}}
}

\begin{document}

\title{Stacking the Deck: Gambling on a Light QCD Axion}
\author{Abhishek Banerjee\orcidA{}\,}
\email{abanerj4@umd.edu}
\affiliation{\UMD}
\author{Manuel A. Buen-Abad\orcidB{}\,}
\email{buenabad@umd.edu}
\affiliation{\UMD}
\author{Anson Hook\orcidC{}\,}
\email{hook@umd.edu}
\affiliation{\UMD}
\date{\today}

\begin{abstract}
We consider axions lighter than what their QCD couplings might otherwise suggest.
Starting with a $\mathbb{Z}_N$-axion, we introduce a small explicit $\mathbb{Z}_N$ symmetry-breaking coupling between the Standard Model Higgs boson and a reheaton.
This small explicit breaking allows us to populate a large portion of the light axion $m_a$--$f_a$ plane, removes the $1/N$ tuning in the $\mathbb{Z}_N$-axion, and explains why only our sector was reheated.
Due to finite temperature effects, axions of this sort undergo either ``rigged" misalignment, where the axion misalignment angle is effectively $\pi$ regardless of its initial value; or ``shuffled" misalignment, where the initial angle is effectively randomized.
\end{abstract}

\maketitle

\section{Introduction} 

One of the intriguing properties of the Standard Model (SM) is the smallness of the neutron electric dipole moment (EDM).  This puzzling fact, 
sometimes called the strong CP problem, begs for an explanation.
A very popular solution to this problem is the QCD axion \cite{Peccei:1977ur,Peccei:1977hh,Weinberg:1977ma,Wilczek:1977pj,Zhitnitsky:1980tq,Dine:1981rt,Kim:1979if,Shifman:1979if}.
The QCD axion dynamically relaxes the neutron EDM to zero, thereby addressing the strong CP problem; at the same time the axion can be cold dark matter (DM), economically solving two problems at the same time \cite{Preskill:1982cy,Dine:1982ah,Abbott:1982af}.

The QCD axion is highly predictive.
The axion mass is uniquely determined by its coupling to QCD, $f_a$, which yield the well-known QCD axion line.
As a popular DM candidate, there are a large number of experiments looking for the QCD axion (see {\it e.g.}~\cite{Ringwald:2024uds,AxionLimits,ParticleDataGroup:2024cfk} and references therein for a comprehensive list). 
For a fixed mass, $m_a$, any given experiment will first exclude couplings stronger than the QCD axion before, hopefully, working its way to the QCD axion line itself.
Because the QCD axion has a fixed QCD coupling for a given mass, it is interesting to ask if the experiments are excluding {\it any} theory on their way to the QCD axion line.
For couplings to fermions and photons, the answer to this question is that the experiments are probing a wide variety of axion-like particle (ALP) models.

The situation is different for the defining coupling of the QCD axion, because the coupling to gluons is also the coupling responsible for producing the mass of the axion.  
Experimentally, this feature leads to a variety of interesting signatures, as finite density effects can be very important~\cite{Hook:2017psm,Balkin:2020dsr,Prabhu:2021zve,Budnik:2020nwz,Balkin:2021wea,Balkin:2021zfd,Noordhuis:2023wid,Balkin:2023xtr,Gomez-Banon:2024oux,Kumamoto:2024wjd,Balkin:2022qer,Banerjee:2025dlo}.
Theoretically, to get an axion with the same mass but larger coupling or, equivalently, the same coupling but a smaller mass, it becomes necessary to cancel the calculable QCD contribution to the mass of the axion.
There is only one theory that achieves this without fine tuning: the $\mathbb{Z}_N$-axion~\cite{Hook:2018jle}.

The $\mathbb{Z}_N$-axion invokes several decoupled copies of the Standard Model (SM).
By having the axion non-linearly realize a $\mathbb{Z}_N$ exchange symmetry, there is an exponential cancellation between the various contributions to the axion mass, making it much lighter than in the canonical QCD case.

There are, however, several features and/or issues with the $\mathbb{Z}_N$-axion.
Because $N$ is an integer, the theory cannot accommodate every single combination of small $m_a$ and $f_a$, but only a discrete set of them.
This is good news for falsifiability, as an axion discovery with $m_a$--$f_a$ values outside of this discrete set of combinations would instantly disprove the $\mathbb{Z}_N$ model, but is also a problem as this implies that there is {\it no} theory that naturally populates the entire parameter space.

The next issue is that observations require that only one of the identical $N$ SM copies have a significant cosmic abundance; all other sectors must have a negligible energy density.  Observational constraints on $\Delta N_{\rm eff} \equiv \rho_{\text{dark rad.}}/\rho_{1\nu} \lesssim 0.3$~\cite{Planck:2019nip}  imply that the temperature of any of the $\mathbb{Z}_N$ copies of the SM must be much smaller than that of our own, spontaneously breaking the $\mathbb{Z}_N$ symmetry.
In other words, $\mathbb{Z}_N$-breaking dynamics {\it must} be part of our cosmic history. 
Finally the last issue to address is that, as a consequence of the non-linear realization of the $\mathbb{Z}_N$ symmetry, the $\mathbb{Z}_N$-axion has $N$ minima, only one of which solves the strong CP problem for our sector (``our" sector being defined as the sector that was reheated).
Thus, solving our strong CP problem requires the axion to land in the correct minimum, which constitutes a $1/N$ accident.

In this paper, we seek to address all of the previous issues with the $\mathbb{Z}_N$-axion.
Our approach is to introduce an explicit, but very small, soft $\mathbb{Z}_N$ symmetry-breaking coupling between the SM Higgs and the reheaton, the particle~\footnote{This could be the inflaton itself, or another particle.} whose decay reheats our sector.
This explicit breaking allows our copy of the SM to be reheated while all the other copies remain cold.
Because of this soft symmetry-breaking coupling, our Higgs boson has a minutely different mass from that of the other Higgses copies, resulting in an additional contribution to the axion mass that is aligned with $\overline \theta = 0$ for our sector.
With the appropriate sign, this additional contribution removes the $N$ vacua degeneracy in favor of a single vacua with $\overline \theta = 0$, solving the $1/N$ tuning problem.
Finally, because there is an additional, extremely small parameter in the reheaton-Higgs potential, the axion mass $m_a$ and $f_a$ are decoupled from each other.
As a result, most of the light axion region in the $m_a$--$f_a$ plane can be populated.
Finally, the global shape of the axion potential is just the canonical QCD axion potential re-scaled to smaller masses.  
This distinction is important when it comes to neutron star constraints, as it is not clear if the bounds of Refs.~\cite{Gomez-Banon:2024oux,Kumamoto:2024wjd} apply to the $\mathbb{Z}_N$ axion, but do apply to the model we present in this paper.

The cosmology of our light axions is slightly different from that of the $\mathbb{Z}_N$ axion~\cite{DiLuzio:2021gos,Co:2024bme}.
As we discuss below, there are two main regions of the axion parameter space, which correspond to two different misalignment regimes.
In the first regime, which we dub {\it rigged misalignment}, there is an early period of oscillations around $\theta = \pi$, effectively ``rigging'' the axion initial conditions to this specific value, regardless of what they might have been before.
As a result of this ``rigging" of the axion's initial conditions, for this parameter space region, the correct dark matter abundance is obtained for a fixed line.
In the second regime, which we call {\it shuffled misalignment}, there is also an early era of oscillations around $\theta = \pi$; however this is followed by a period during which most of the axion energy is kinetic, resulting in a randomized effective axion misalignment angle.
This randomization is sensitive to percent level changes in the initial condition, and thus it can be said that it re-``shuffles'' them ``fairly".
Much like in the canonical misalignment mechanism scenario, our shuffled misalignment regime can result in axion dark matter having the correct abundance, granted some luck in its initial conditions.

In the rest of this paper we flesh out this idea in more detail.
In Sec.~\ref{sec:model}, we discuss the model. In Sec.~\ref{sec:cosmology} we discuss the cosmological evolution of our model. 
In Sec.~\ref{sec:axion_dark_matter}, we discuss its implications for dark matter.
Finally, we conclude in Sec.~\ref{sec:conclusion}.

\section{Model}\label{sec:model}

In this section we present our model, which is a modification of the $\mathbb{Z}_N$ axion.
We begin with a review of the original model, after which we provide a description of our changes, as well as their effects on the axion potential.

\subsection{The $\mathbb{Z}_N$-axion}

The starting point of the $\mathbb{Z}_N$-axion is $N$ identical copies of the SM, related to each other by a $\mathbb{Z}_N$ exchange symmetry that is also non-linearly realized on the axion~\cite{Hook:2018jle}:
\bea\label{eq:ZN_definition}
\!\!\!\!\!\!\!\!\!\!\!\!
\mathbb{Z}_N: \,\,\cL_{{\rm SM},k}\to \cL_{{\rm SM},\,k+1}\,\, \text{ and }\,\, \frac{a}{f_a}\to \frac{a}{f_a} + \frac{2\pi}{N}\,\,.
\eea
The Lagrangian terms relevant for the axion are:
\bea\label{eq:Ltot}
\cL \supset \frac{\alpha_s}{8\pi}\sum_{k=0}^{N-1} \left(\frac{a}{f_a}+\overline{\theta} + \frac{2\pi k}{N}\right) \bl( G\tilde{G} \br)_k \,\,,
\eea
where $\alpha_s$ is the strong coupling constant. 
The $k$-th SM copy has a strong CP angle given by $\overline{\theta}_k \equiv \overline{\theta} + 2\pi k/N$; and without loss of generality we take the $k=0$ SM copy to be our sector.

The axion coupling to each of the $( G\tilde{G} )_k$ generates an axion potential.
To leading order in chiral perturbation theory ($\chi$PT), the canonical QCD axion potential generated by this coupling is
\beq
\label{eq:QCD_axion_potential}
V\bl(\theta + \overline{\theta}\br) = -\frac{m_\pi^2 f_\pi^2}{1+z} \sqrt{1+z^2 + 2z  \cos \bl(\theta+\overline{\theta}\br)} \,,
\eeq
where we define $\theta\equiv a/f_a$, and $m_\pi \approx 135\MeV$, $f_\pi\approx 93\MeV$ and $z=m_u/m_d=0.47$ are the pion mass, pion decay constant, and the up-to-down quark mass ratio respectively~\cite{ParticleDataGroup:2024cfk}.  
This potential gives the QCD axion a mass
\bea
\mqcds = \frac{z}{(1+z)^2}\frac{m_\pi^2 f_\pi^2}{f_a^2} \,.
\eea

The total axion potential generated from the couplings in Eq.~\eqref{eq:Ltot} can be written as
\bea\label{eq:ZN_Vtot}
V_N(\theta) && = \sum_{k=0}^{N-1} V \bl( 
\theta+ \overline\theta_k \br)\, \\ \xrightarrow[]{N \gg 1} && \frac{m_\pi^2 f_\pi^2}{\sqrt{N \pi}} \sqrt{\frac{1-z}{1+z}} 
\left ( - z \right )^N \cos N \theta + \cO(z^{2N}) \nn \\
& &\equiv (-1)^N \frac{(m^2_a)_N f^2_a}{N^2}\cos N\theta \,,
\eea
For notational convenience, we have used a shift in the axion field in order to set $\overline{\theta}=0$, and we have included the large $N$ limit of the potential~\cite{DiLuzio:2021pxd}.
Furthermore, we have dropped all constant terms (\ie, those not depending on the axion field itself), and in the last equality we ignored the $\cO(z^{2N})$ terms. 
Therefore, for large $N$ the axion mass is suppressed with respect to the canonical case, namely
\bea\label{eq:epsN}
\!\!\!\!\!\eps_N\equiv \frac{(m^2_a)_N}{(m^2_a)_{\rm QCD}} \simeq \sqrt{\frac{1-z^2}{\pi}}(1+z) \ N^{3/2} \, z^{N-1} \,.
\eea
This exponential suppression is due to the analytic nature of the axion potential, as well as to the fact that Eq.~\eqref{eq:ZN_Vtot} is a Riemann sum~\cite{Hook:2018jle,DiLuzio:2021pxd}.

\vspace{12pt}

\subsection{Our modification}
\label{sec:idea}

As mentioned before, we introduce a soft, $\mathbb{Z}_N$ symmetry-breaking coupling between the reheaton, denoted by $\phi$, and the Higgs, $H$.
Namely, we consider the coupling 
\bea
\label{eq:g_coupling}
\mathcal{L} \supset -\phi \sum_{k=0}^{N-1} g_k |H_k|^2\, - V_\phi(\phi),
\eea
where $g_k$ is a dimension-one coupling.
A coupling of this form can be enforced by requiring an additional $\mathbb{Z}_2$-symmetry under which both $\phi$ and $g_k$ are odd.

For $m_\phi \gtrsim 2m_h$, the reheaton decays into a pair of Higgses $H_k$, thereby reheating the $k$-th SM sector.
If all of the $g_k$ were equal, then the $\mathbb{Z}_N$ symmetry would be restored, and all sectors would be equally reheated.
For simplicity, let us assume maximal $\mathbb{Z}_N$ symmetry-breaking and, without loss of generality, choose $g \equiv g_0 \neq 0$, and $g_{k \neq 0} = 0$.
This guarantees that, upon decaying, the reheaton reheats only one sector, ours.\footnote{We would like to reiterate that what makes our sector {\it ours} is the fact that it is the only one currently present in the Universe. Thus, for a maximally $\mathbb{Z}_N$-breaking reheaton-Higgs sector (where the reheaton couples to only one of the Higgses), we can always call the reheated sector ours. Choosing $k=0$ as its label is entirely arbitrary.}

\paragraph{The Higgs VEV:} Aside from causing asymmetric reheating, Eq.~\eqref{eq:g_coupling} also induces a small shift to the Higgs VEV.

To see exactly how this works, let us begin by writing down the potential for the SM ($k=0$) Higgs and the $\phi$ reheaton:
\bea\label{eq:2scalars}
V(\phi, H) &&= -(\mu^2- g\phi)(H^\dagger H)+ \lambda_H (H^\dagger H)^2 \nn\\
&&+ \frac{1}{2} m_\phi^2 \phi^2 \pm g\frac{\Lnp^2}{16\pi^2} \phi\,.
\eea
The last term is an estimate for the size of the linear term allowed by symmetries.
This estimate arises from closing the Higgs loop in the $\phi$ self-energy diagram, and it involves a UV cutoff $\Lnp$ where the effective field theory breaks down.
Throughout this paper, we take $g$ to be the smallest scale in the problem, $g \ll m_\phi \lesssim \Lnp$.
A typical value of $g$ is $g \sim 10^{-7} m_\phi$.
This choice of $g$ is technically natural since, in the limit of $g\to 0$, the higher-spin symmetries for $\phi$~\cite{Hook:2023pba} are restored. 
We have neglected the $\phi^3$, $\phi^4$ and $\phi^2|H|^2$ couplings in our analysis, as for our parameter choices these interactions are small. Their expected sizes are $\gtrsim \mathcal{O}(g^3/16\pi^2 m_h^2)$, $\mathcal{O}(g^4/16\pi^2 m_h^4)$, and $\mathcal{O}(g^2/m_h^2)$ respectively.

After electroweak symmetry breaking, both $\phi$ and $H$ obtain vacuum expectation values (VEVs) which can be written as, 
\begin{widetext}
\beq\label{eq:field_vevs}
    \vev{\phi} = \phi_0 = - \frac{g}{m_\phi^2}\left(\frac{v_0^2}{2}\pm\frac{\Lnp^2}{16\pi^2}\right) \ , \quad
    \vev{H^2} = \frac{v_0^2}{2}  \simeq \frac{v_{k \neq 0}^2}{2} \left[1+ \frac{g^2}{2 \lambda_H m_\phi^2}\bl(1\pm  \frac{\Lnp^2}{8\pi^2 v_{k \neq 0}^2}\br)\right]\,,
\eeq
\end{widetext}
where $v_0 = 246\,\GeV$ is the VEV of our sector's Higgs, and $v_{k \neq 0}$ as the VEV of the other $H_{k\neq0}$ Higgses.
We have replaced $\mu^2/\lambda_H = v_{k \neq 0}^2$ and, in the last equality, performed a Taylor expansion for $g^2/2 \lambda_H m_\phi^2 \ll 1$.
There is a runaway behavior of the potential for $\lambda_H \lesssim g^2/(2 m_\phi^2)$, which can be cured by introducing higher order terms such as $\phi^2 |H|^2$, $\phi^4$ etc.
For our purposes, we take instead $\lambda_H \gg g^2/(2 m_\phi^2)$, and obtain from Eq.~\eqref{eq:field_vevs} 
\bea
\label{eq:deltav_vev}
\!\!\!\!\!\!\!\!\!\!\!\!
\frac{\Delta v}{v_0} \equiv \frac{v_0 - v_{k \neq 0}}{v_0} 
& \simeq & \frac{g^2}{4 \lambda_H m_\phi^2} \bl(1 \pm \frac{\Lnp^2}{8\pi^2 v_0^2}\br).
\eea
Hence we find that, due to the interaction between the SM Higgs and the reheaton, the Higgs VEV of our sector is changed with respect to every other $N-1$ sector with $k \neq 0$.
As will become clear in the subsequent sections, for our model to solve the strong CP problem we need the VEV of our sector's Higgs to be larger than the VEV of the other Higgses; therefore we choose $\pm \rightarrow +$.

The end result of this modification to the $\mathbb{Z}_N$-axion is a standard Higgs portal, which has been extensively studied in the literature (see {\it e.g.}~\cite{OConnell:2006rsp,Piazza:2010ye, Fuchs:2020cmm} and references therein). 
While we do not review the Higgs portal scenario in this paper, we want to highlight an important point.
Demanding that one‐loop radiative corrections be no larger than the tree-level values for $m^2_\phi$, $m_h^2$, and $\lambda_H$, we obtain a constraint on $g$ as $g/m_\phi\lesssim {\rm min}\left[4\pi m_h/m_\phi,\sqrt{\lambda_H}\right]$ (ignoring the cut-off dependent logarithmic factor, which is $\sim\cO(10)$ for $\Lambda_{\rm NP}=100\TeV$).
This constraint is easily satisfied for the sufficiently small values of $g$ in which we are interested. 
\\

\paragraph{The Axion Potential:} When the Higgs VEVs of the $N$ sectors are identical, the axion potential is given by Eq.~\eqref{eq:ZN_Vtot}.  
When the Higgs VEV of our $k=0$ sector is changed, the axion potential is changed to
\bea
\label{eq:potential_T_0}
\overline{V}\left(\theta\right) &=& (1+\ephi) V\bl(\theta\br)+\sum_{k=1}^{N-1} V\bl(\theta+ \frac{2\pi k}{N}\br) \nn\\ 
&=&  \ephi V(\theta) + V_N(\theta)\,.
\eea
$\ephi\sim \Delta v/v_0$ comes from a Taylor series of Eq.~\eqref{eq:QCD_axion_potential} with a small change in the Higgs VEV and is explicitly  derived in Eq.~\eqref{eq:ephi_def}.
Around $\theta=0$, the axion mass can be written as
\beq
    \frac{m_a^2}{(m_a^2)_{\rm QCD}} =   (-1)^{N+1} \epsilon_N + \ephi \ll 1  \,.
\eeq
Thus, unlike in the original $\mathbb{Z}_N$-model, as long as $1 \gg \ephi \gtrsim \eps_N$, our model yields a light QCD axion {\it for any $N$} (even or odd). 
Since the $\phi$-induced contribution to the potential is only minimized at $\theta = 0$, the strong CP problem of {\it our} sector is {\it always} solved, without the need for a linear $1/N$ tuning. 
Finally, as $m_a\simeq \sqrt{\ephi} \mqcd$, in our model, the axion cannot be made arbitrarily light, unlike in the $\mZ_N$ case.
As will be clearer when cosmology is discussed in the next section, $\ephi$ is connected to the reheating temperature and thus the minimum $\ephi$ our model permits is $\mathcal{O}(10^{-24})$ (for $\Trh\sim 10\MeV$).
Thus, the axion in our model can be up-to $\mathcal{O}(10^{-12})$ lighter than the canonical QCD axion (see Fig.~\ref{Fig:ephi_trh}).

The last step in our axion potential calculation is to derive $\epsilon_\phi$.  The change in the Higgs VEV shifts both the light quark masses and the confinement scale (through the running of $\alpha_s$). 
Using the one-loop $\beta$-function, we find that the change in the QCD scale due to the shift in the Higgs VEV as 
$\Lambda_{\rm QCD, 0}/\Lqcd = (v_0/v_{k \neq 0})^{2/9}$ for three light quark flavors~\cite{Berezhiani:2000gh}. 
From Eq.~\eqref{eq:QCD_axion_potential} and using $m_\pi^2 f_\pi^2\propto m_q\Lambda_{\rm QCD}^3\propto v^{5/3}$ with $q=u,d$, we find
\bea\label{eq:ephi_def}
\ephi &\equiv & \frac{\Delta \bl ( m_\pi^2 f_\pi^2 \br)}{m_\pi^2 f_\pi^2} = \frac{5}{3} \frac{\Delta v}{v_0} \nn\\ 
&\simeq & \frac{ 5g^2}{12\lambda_H m_\phi^2} \bl(1+\frac{\Lnp^2}{8\pi^2v_0^2}\br) \,. 
\eea
We also use Eq.~(\ref{eq:deltav_vev}) in deriving the above formula.

\section{Axion Cosmology}\label{sec:cosmology}
In this section, we discuss the cosmological implications of our model. We begin by analyzing the reheating phase, and show how the presence of a light QCD axion is intertwined with the reheating temperature. 
Subsequently, we investigate the viability of axion dark matter in our model. 
\subsection{Thermal History}\label{subsec:thermal}
To obtain the reheating temperature of our sector, we calculate the reheaton decay width to the SM Higgs $\phi \to H^{\dagger}H$ as 
\bea
\Gamma_{\phi} \simeq 
\frac{g^2}{8\pi\, m_\phi}\,,
\eea
in the high-energy limit. 
For simplicity, we will set 
$(m_\phi)_{\rm min}\sim 500\GeV$ as the smallest $\phi$ mass for which the reheaton can still efficiently decay into a Higgs pair and thereby reheat the Universe.

Our sector is reheated to a temperature $\Trh$ defined as $\Gamma_{\phi} = H(\Trh) \equiv\Hrh =\sqrt{\pi^2 g_*(\Trh) / 90} \, \Trh^2 / \Mpl$ (with $\Mpl=2.4\times 10^{18}\GeV$ being the reduced Planck mass) while the other $N-1$ sectors remain cold.
Thus, any $\Delta N_{\rm eff}$ constraints are readily avoided by the simple fact that the $\phi$ reheaton decays only to the $k=0$ sector.
The reheating temperature is therefore
\bea
\!\!\!\!\!\!\!\!
\label{eq:Trh_delta}
\frac{\Trh}{1\TeV}
\simeq \bl( \frac{g/m_\phi}{2 \times 10^{-7}} \br) \sqrt{\frac{m_\phi}{1\TeV}} \bl( \frac{106.75}{g_*(\Trh)} \br)^{1/4} .
\eea
Other decay channels of $\phi$ to the SM are sub-dominant as they are suppressed by the $h-\phi$ mixing angle, $\sin^2\theta_{h\phi}\sim (g v_0/m_\phi^2)^2\ll 1$. 
Since the coupling that shifts the Higgs VEV also leads to reheating, by combining Eq.~\eqref{eq:Trh_delta} and Eq.~\eqref{eq:ephi_def}, we obtain 
\begin{align}
\label{eq:ephi_Trh}
\ephi \simeq \frac{10 \pi}{3 \lambda_H} \left(1+\frac{ \Lnp^2}{8\pi^2v_{0}^2}\right)  \left(\frac{\Hrh}{m_\phi}\right)
\,.
\end{align}
Given that $\epsphi\propto \Trh^2$, a smaller value of $\ephi$ is tied to a lower reheating temperature.  To emphasize this, let us estimate the smallest $\epsphi$ can be.

\begin{figure}[t!]
    \centering
    \includegraphics[width=\linewidth]{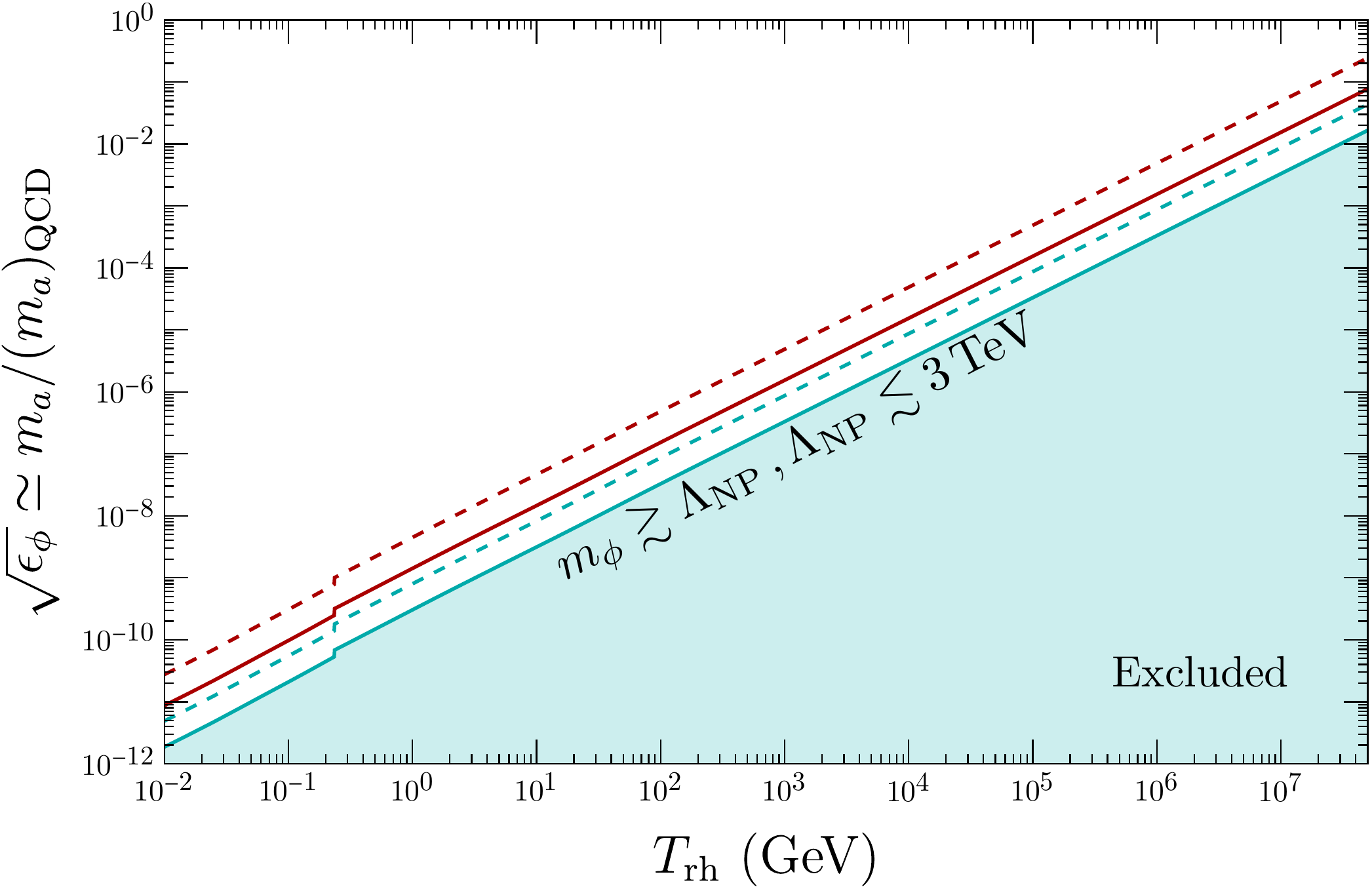}
    \captionsetup{justification=justified}

    \caption{
    Plot of the lightness of the axion mass {\it i.e.} $\sqrt{\ephi}$ as a function of the reheating temperature $\Trh$, for different choices of $\Lnp$ and $m_\phi$. 
    The red and turquoise lines show the parameter space for $\Lnp=100\TeV$ and $3\TeV$ respectively.
    The solid lines indicate the minimum value of $\ephi$ that can be obtained for a given $\Lnp$ with $m_\phi/\Lnp=1\,$ whereas the dashed red and dashed turquoise lines are for $m_\phi=10\TeV$ and $0.5\TeV$ respectively. 
    The values of $\ephi$ in the turquoise shaded excluded region require $m_\phi\gtrsim \Lnp$ for $\Lnp=3\TeV$, where the validity of the EFT breaks down. The sudden drop in the values of $\ephi$ around $\Trh\sim 0.2\GeV$, is due to the change in $g_*(T)$ close to the QCD phase transition.  
    }    
    \label{Fig:ephi_trh}
\end{figure}

From Eq.~\eqref{eq:ephi_Trh}, we have $\ephi\propto \Lnp^2 \Trh^2/m_\phi$.
Therefore, in order to obtain the smallest $\epsphi$, we need $\Lnp$ and $\Trh$ to be as small as possible, while on the other hand keeping $m_\phi$ as large as possible.
We conservatively take $(\Lnp)_{\rm min}=4\pi v_0\approx3\TeV$.
The largest allowed $\phi$-mass, for which the effective field theory (EFT) description does not break down, is of order $\sim \cO(\Lnp)$.
Thus, taking $\Lnp = m_\phi = 3\TeV$ and a very low reheating temperature of $\Trh=10\MeV$, we find that the smallest value of $\ephi$ allowed by our model is $\ephi \approx 3\times 10^{-24}$.
For comparison, another reasonable benchmark point in parameter space is $\Lnp=100\TeV$, $m_\phi=10\TeV$, and $\Trh=1\TeV$, which yields the larger value of $\ephi \approx  2\times 10^{-11}$.
In Fig.~\ref{Fig:ephi_trh} we plot this minimum $\ephi$ for a given cut-off of the theory, as a function of $\Trh$. 
The turquoise and the red lines represent $\Lnp=3\TeV$ and $100\TeV$ respectively, with the solid and dashed textures corresponding to different choices of $m_\phi$.
The sudden jumps in $\ephi$ are caused by the abrupt change in the degrees of freedom $g_*$ of the thermal bath around the QCD phase transition.

For the other $N-1$ sectors not to thermalize with the SM bath via a tree level $a$ exchange (processes such as $(G\tilde G)_{k=0} \to (G\tilde G)_{k\neq 0}$), its associated rate must satisfy $\Gamma_{0\to k}\sim T^5 (\alpha_s/8\pi f_a)^4\lesssim \Hrh $, resulting in the following bound:
\bea\label{eq:fa_cond1}
f_a \gtrsim 18 \TeV\, \left(\frac{\alpha_s}{0.09}\right)\left(\frac{\Trh}{1\TeV}\right)^{3/4}\,.
\eea
More conservatively, we impose an even stronger condition by demanding that the axion not be thermalized (\eg, via $(G\tilde G)_{k=0}\to a \, G_{k=0}$).
The rate associated with this process must obey $\Gamma_{0\to a}\sim T^3 (g_s \alpha_s/8\pi f_a)^2\lesssim\Hrh$, which results in~\cite{Masso:2002np}
\bea\label{eq:fa_cond2}
 f_a \gtrsim 4\times 10^8\GeV\, \left(\frac{\alpha_s}{0.09}\right)^{3/2}\left(\frac{\Trh}{1\TeV}\right)^{1/2}\,.
\eea
For $\Trh\lesssim T_c$, we consider axion thermalization through the pions and nucleons of the $k=0$-th sector ({\it e.g.} via $\pi\pi \to a\pi$ and/or $a N\to N\pi$ processes).  Neglecting phase space factors, which only loosen the constraint, Ref.~\cite{Hannestad:2005df} obtains 
\bea
\label{eq:fa_cond3}
f_a\gtrsim 2\times 10^{7} \GeV \left(\frac{T}{0.1\GeV}\right)^{3/2}\!.
\eea
for the axion not to be thermalised. 
Throughout this paper, we take $f_a$ such that Eqs. (\ref{eq:fa_cond1}) and (\ref{eq:fa_cond2}) or Eq.~\eqref{eq:fa_cond3} are satisfied.
\\

\subsection{Axion Dark Matter} \label{sec:axion_dark_matter}
In this subsection, we explore the axion cosmology of our model.  We will pay special attention to the regions of parameter space where the axion can also play the roll of dark matter.
While all of our plots are obtained numerically, the majority of this section will instead be dedicated to obtaining an analytical understanding of the dynamics.

Since only the $k=0$ sector is reheated, the temperature ($T$) dependence of the axion potential originates from the $0$-th sector. 
The temperature-dependent axion potential can be written as 
\bea\label{eq:potential_T}
\overline{V}(T,\theta) & \simeq & (1+\ephi)F(T) V(\theta) + \sum_{k=1}^{N-1} V\bl(\theta+ \frac{2\pi k}{N}\br)\nn\\
& = & V(\theta) \bl[ (1 + \ephi) F(T) - 1 \br] + V_N(\theta)
\eea
where $F(T)$ captures the temperature dependence of the topological susceptibility.  Roughly speaking, $F(T)$ has the behavior
$F(T)=(T_c/T)^8$ for $T\gg T_c$, and $F(T)=1$ for $T\ll T_c$, with $T_c\simeq 150\MeV$ being the QCD phase transition temperature~\cite{Borsanyi:2016ksw,Petreczky:2012rq,GrillidiCortona:2015jxo}\footnote{
The exact functional form of $F(T)$ close to the phase transition temperature is an active area of research~\cite{Borsanyi:2016ksw,Petreczky:2012rq,GrillidiCortona:2015jxo}. 
In our numerical calculations, we interpolate between the high temperature and low temperature behaviors by writing $F(T) = 1/[1+(T/T_c)^{8\beta})]^{1/\beta}$, for $\beta>0$. 
A sharper (smooth) phase transition can be achieved for $\beta \gg 1$ ($\beta \ll 1$). 
For our qualitative estimates, the exact value of $\beta$ bears no significant change. 
In all our plots below we take $\beta=12$.}.
The  axion potential can be written explicitly as
\begin{widetext}
\bea
\frac{\overline{V}(T,\theta)}{(m_a^2)_{\rm QCD} f_a^2}& = &  (-1)^N \frac{\eps_N}{N^2}\cos(N\theta) + \frac{(1+z)}{z} \left[ 1 - (1+\ephi) F(T) \right]\sqrt{1+ z^2+2 z\cos\theta} + \text{ const.}
\eea
\end{widetext}
At high temperatures, $F(T) \ll 1$ and the above potential is minimized at $\theta=\pi$. 
The axion mass around this high-temperature minimum can be written as
$
m^\pi_a \equiv m_a(T \gg T_c) \simeq \mqcd\sqrt{(1+z)/(1-z)}\,,
$
which for $z \approx 0.5$ is larger than the canonical QCD axion by a factor of $\sim \sqrt{3}$. 
On the other hand, for $T\ll T_c$, $F(T) \simeq 1$ and the potential is minimized at $\theta=0$, with the axion mass suppressed by $\ephi$, as discussed previously.
Therefore, unlike in the case of the canonical QCD axion, in our model the axion was parametrically several orders of magnitude {\it heavier} in the early universe than it is today (by a factor of about $\sqrt{3/\epsilon_\phi}$ for $\epsphi \ll \eps_N$); additionally, in the early Universe the minimum of the high-temperature potential was located at $\theta = \pi$ instead of $\theta = 0$.
This represents a significant departure from the standard misalignment mechanism, where the potential is flat at very high temperatures and a significant tuning is required to obtain an initial angle $\sim \pi$ ({\it c.f.} refs.~\cite{DiLuzio:2021pxd,DiLuzio:2021gos}).

It is now evident that, for large reheating temperature $\Trh\gtrsim T_c$, the axion in our model can undergo a two-stage misalignment mechanism: first evolving towards the high-temperature minimum at $\theta = \pi$, before eventually turning around and evolving towards the low-temperature global minimum at $\theta = 0$. 
Since $m^\pi_a \approx  \sqrt{3} \, \mqcd$, the two oscillatory stages occur as long as $\mqcd \gtrsim H(T_c) \equiv H_c$, irrespective of the value of $\ephi$.
Numerically, we find that the axion undergoes one full oscillation around this minimum if $f_a \lesssim (f_a)_{\rm crit} \approx 5\times 10^{16}\GeV$, corresponding to $\mqcd \approx 8 H(T_c)$. 
The time $t_{\rm osc,\pi}$, at which this first era of oscillations begins, is obtained by solving $3H (t_{\rm osc,\pi}) = {\rm min}[\mapi, 3 \Hrh]$, and the corresponding amplitude of the axion field can be written as
\begin{align}
\label{eq:oscillation_pi}
\pi-\theta(t) \simeq \left|\pi-\theta_i\right| \left(\frac{\aoscpi}{a}\right)^{3/2}\cos[m^\pi_a (t-t_{\rm osc,\pi})]\,,
\end{align}
where $\theta_i \equiv \theta(t_i)$ is the initial misalignment angle of the axion field at some initial time $t_i$, which we take to be the time of reheating $t_i=\trh$.
Thus, we have $\theta_i = \theta_{\rm rh}.$\footnote{Henceforth, we denote $\theta_x \equiv \theta(t_x)$ the value of the axion field $\theta(t)$ at some time of interest $t_x$. The same is true of the axion field velocity, namely $\dot\theta_x \equiv \dot\theta(t_x)$.}
As $T$ nears the QCD phase transition temperature, $F(T)\to 1$ and the temperature-dependent part of the axion potential vanishes. 
The potential develops a new minimum at $\theta=0$, furthermore becoming shallower as its amplitude gets multiplied by $\ephi$.
At this point the axion mass changes from $\mapi \approx \sqrt{3} \mqcd$ to $\mqcd \sqrt{\ephi} \ll \mqcd$.\footnote{The axion potential, and thus the axion mass, vanishes when $T_c/T = [(1+\epsphi)^\beta-1]^{-1/8\beta}\simeq (\beta\epsphi)^{-1/8\beta}$.
Since $(\beta\epsphi)^{-1/8\beta}\gtrsim 1$ for $\epsphi\lesssim 1$, the amplitude of the axion potential starts to decrease as $T\to T_c$, and it vanishes at some temperature just after the phase transition which is dictated by $\beta$. 
The next phase of the axion evolution depends on the zero temperature axion mass, \ie~$m_a=\mqcd \sqrt{\epsphi}$, and on $H(T_c)$. 
We find for $\beta=12$, the temperature at which axion mass vanishes is $\sim 100\MeV$ for $\ephi=10^{-20}$, and it gets closer to $T_c$ as we increase $\ephi$.}
At the time $t_c$ of the QCD phase transition, the axion will be very close to $\pi$ and will have some velocity leftover from the first era of oscillations, parametrically given by $|\dot\theta_c|\sim  \mapi |\pi-\theta_c|$.

The next stage of axion evolution depends on how its zero temperature mass compares to the Hubble scale at $T_c$.  In the first case, which we call {\it shuffled misalignment}, $H_c\gtrsim m_a=\mqcd\sqrt{\ephi}$.  During shuffled misalignment, the axion field value is effectively randomized before it starts oscillating again.  In the second case, which we call {\it rigged misalignment}, $ m_a=\mqcd\sqrt{\ephi}  \gtrsim H_c$.  During rigged misalignment, the axion field value settles close $\theta \approx \pi$ irrespective of the initial field value, and its subsequent evolution is determined by its leftover velocity.
\\

\paragraph{Shuffled misalignment:}
Let us first focus on the $H_c\gtrsim m_a=\mqcd\sqrt{\ephi}$ case, for which the axion potential is extremely flat around the phase transition.
Because of this, the axion is kinetic energy-dominated so that ${\ddot\theta}+ 3 H\dot\theta\simeq 0 \Rightarrow \dot\theta \propto a^{-3}$.  
The kinetic energy dominated phase continues until $H\sim \mqcd \sqrt{\ephi}$, at which point the axion starts to oscillate around the CP preserving minimum.
Thus, in between the QCD phase transition, and the onset of the second oscillatory stage, the field excursion $\Delta \theta\equiv \theta_{\rm osc,0}- \theta_c$, can be written as $\Delta \theta \sim \pm |\pi-\theta_{\rm rh}| \left(\mqcd/H_c\right)^{1/4}$, where $\theta_{\rm osc,0} = \theta(t_{\rm osc,0})$ is the field amplitude at the onset of the second oscillatory era around $\theta = 0$, and $3H(T_{\rm osc,0}) = \mqcd \sqrt{\ephi}$.
The choice in $\pm$ in the above equation depends on the sign of $\dot\theta_c$, which is effectively random due to the fast axion oscillations around $\theta = \pi$.
Because $\mqcd,\Hrh\gg H_c$, $\Delta \theta$ can easily be of order $\mathcal{O}(\pi)$. 
Thus the rolling phase generically makes $\theta_{\rm osc,0}$ very different from $\theta_{\rm rh}$, effectively randomizing (``shuffling'') the misalignment angle in preparation for the second oscillatory phase.
Numerically, we find that initial conditions differing by at least $\cO(1\%)$ are enough to change $\theta_{\rm osc,0}$ by order one.

Once the Hubble scale becomes comparable to $\mqcd \sqrt{\ephi}$, the field starts to oscillate once again, but this time around $\theta=0$. 
The energy density of the axion can be written as
$\rho_a(t) =  m_a^2 f_a^2\theta^2(t)/2$
which can account for present day dark matter energy density when
\begin{align}\label{eq:fa_adm_shuffle}
f_a\simeq 2\times 10^{14}\GeV \left(\frac{\eps_\phi}{2\times 10^{-10}}\right)^{-\frac{1}{6}} \left(\frac{\theta_{\rm osc,0}}{\pi/\sqrt{3}}\right)^{-\frac{4}{3}}\left(\frac{r}{1}\right)^{\frac{2}{3}}\,,
\end{align}
where $r=\Omega_a/\Omega_{\rm DM}$ and $\pi/\sqrt{3}$ serves as a benchmark for the RMS value of an initial amplitude for $\theta$ (in the harmonic limit)~\cite{OHare:2024nmr}.

In the above equation, we have taken $\theta_{\rm osc,0}$ to be a free parameter, as it depends sensitively on all of the other parameters in the theory.
As stated above, even percent-level changes in $\theta_{\rm rh}$ can change $\theta_{\rm osc,0}$ by an order one amount. 
Numerically, this extreme sensitivity comes about at the transition between the two phases, namely when the axion oscillates around $\theta = \pi$, and when the axion becomes kinetic energy-dominated. 
Because even two very similar initial conditions may result in $\cO(1)$ differences in their respective axion field values after the QCD phase transition, we deem this scenario {\it shuffled misalignment}.

\begin{figure}[t]
    \centering
    \begin{subfigure}{0.45\textwidth}
        \centering
        \includegraphics[width=\linewidth]{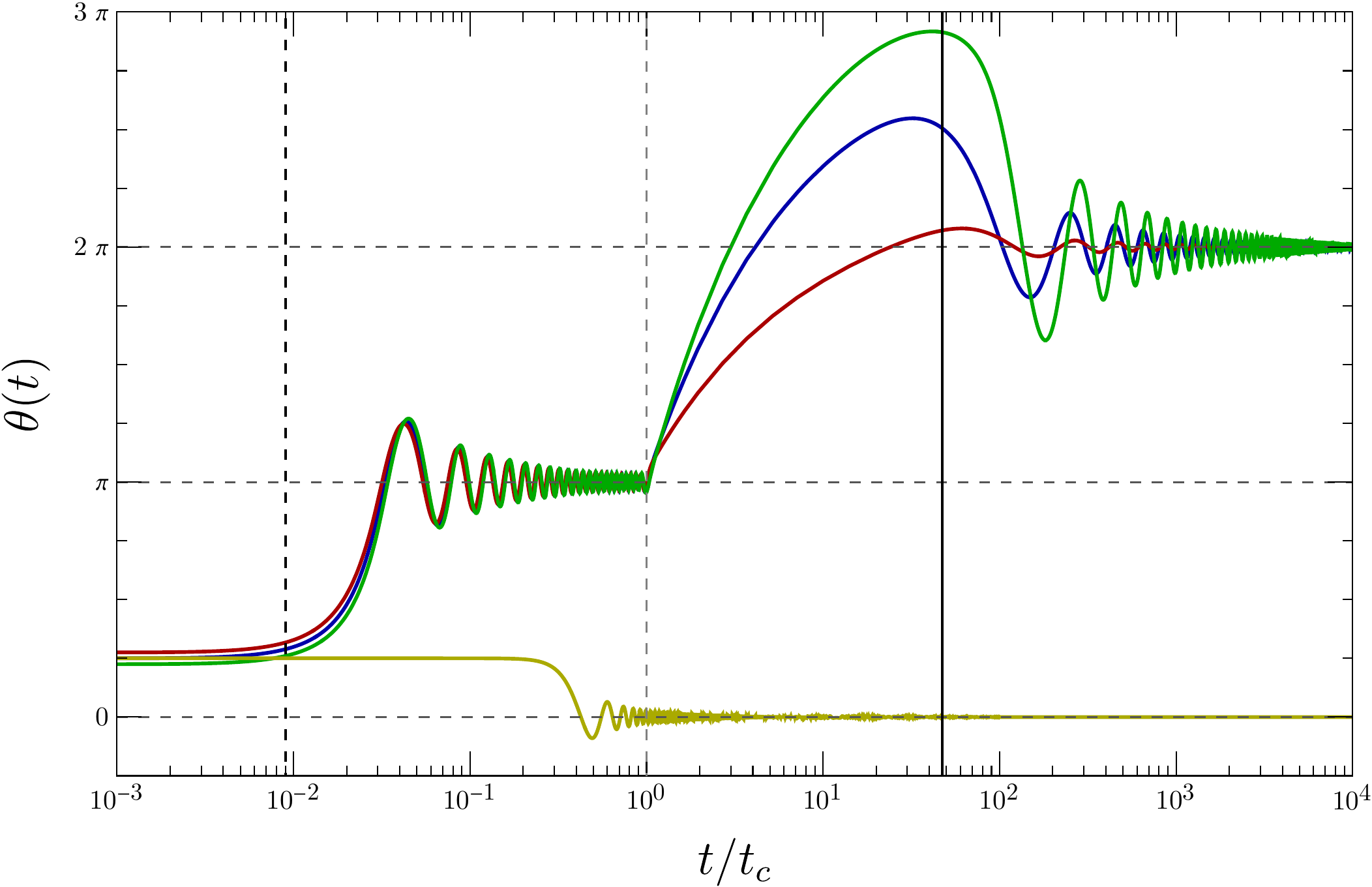}
    \end{subfigure}
    \begin{subfigure}{0.45\textwidth}
        \centering
        \includegraphics[width=\linewidth]{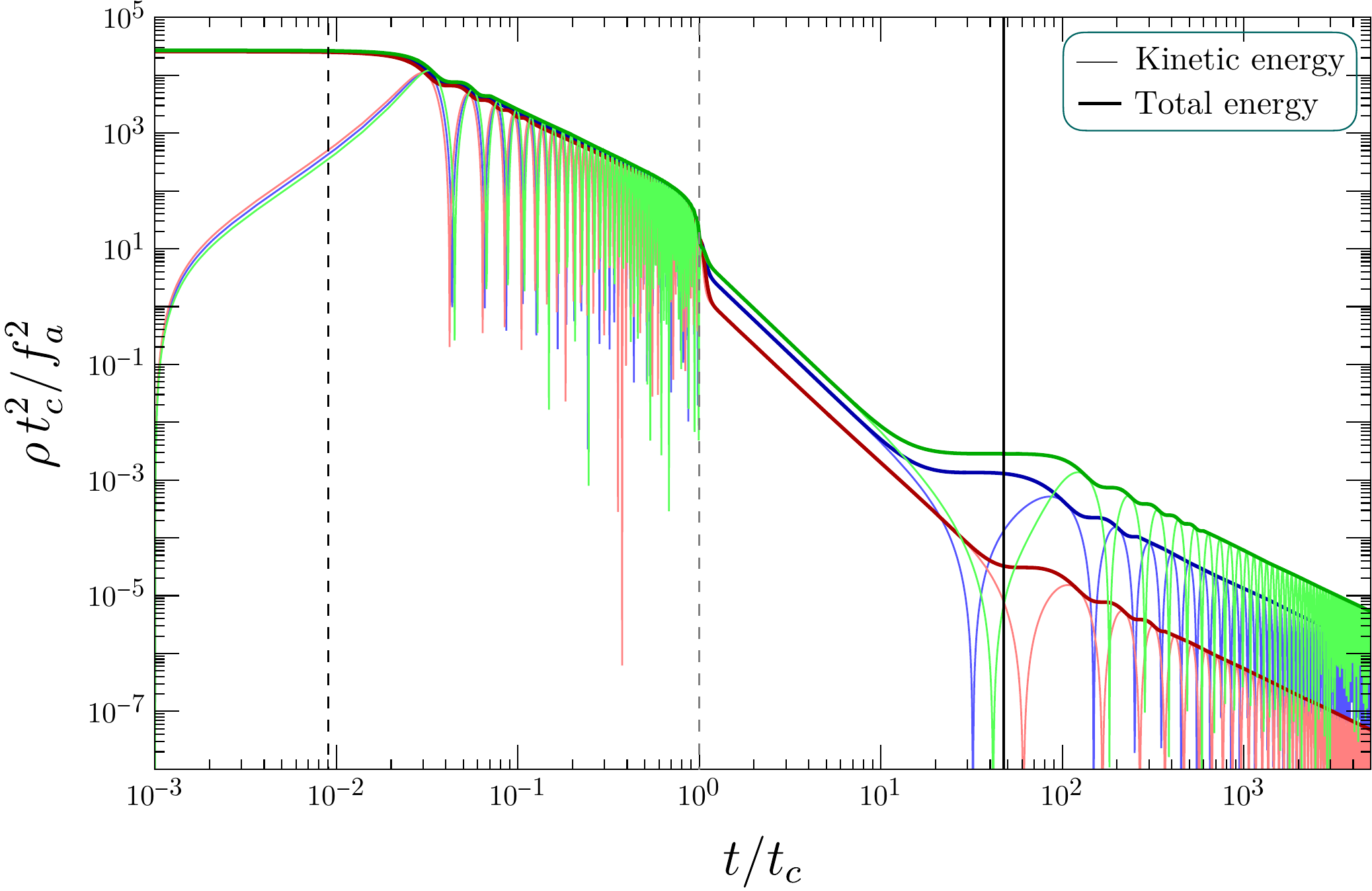}
    \end{subfigure}
    \caption{{\it Shuffled misalignment}. 
    The evolution of the axion angle $\theta$ ({\bf top}) and its normalized energy density $\rho_a \, t_c^2/f_a^2$ ({\bf bottom}) as a function of time $t$, for $f_a\simeq 2\times 10^{15}\GeV$ and $\ephi=10^{-7}$.
    {\it Top:} The evolution of $\theta(t)$ for three very similar initial conditions: $\theta_{\rm rh} = \pi/4$ (blue), and $\theta_{\rm rh} = \pi/4(1\pm10\%)$ (red and green, respectively).
    In yellow we plot the evolution of a canonical QCD axion with same $f_a$ and $\theta_{\rm rh}=\pi/4$.
    The dashed (solid) black line shows $t_{\rm osc,\pi}$ ($t_{\rm osc,0}$), when the axion begins to oscillate around $\theta=\pi$ ($\theta=0$).
    The gray dashed vertical line marks $t_c$.
    Note that, due to the periodicity of the axion potential, the vertical axis in the top panel must be taken to be modulo $2\pi$.
    Therefore the final oscillatory stage, shown to be around $\theta = 2\pi$, can equivalently be interpreted to take place around $\theta=0$.
    {\it Bottom:} The evolution of the energy density of the axion.
    The solid lines correspond to the total energy density, while the thin lines to the kinetic energy term only.
    The colors have the same meaning as in the top panel, and so do the vertical lines.
    }
    \label{fig:case_1a}
\end{figure}

In Fig.~\ref{fig:case_1a} we depict the evolution of the axion field ($\theta$, top panel) and its normalized energy density ($\rho_a \,t_c^2/f_a^2$, bottom panel) in the shuffled misalignment regime as a function of time $t$, for $f_a = 2 \times 10^{15}\GeV$ and $\ephi=10^{-7}$. 
In the top panel, the three lines plot $\theta(t)$ for three very similar initial conditions: $\theta_{\rm rh} = \pi/4$ (blue) and $\theta_{\rm rh} = \pi/4(1\pm 10\%$) (red and green, respectively.
For comparison, we show as well the evolution of a canonical QCD axion with the same $f_a$, and $\theta_{\rm rh} = \pi/4$, in yellow.
There are clearly four distinct periods of evolution for the axion.
In the first, the axion value remains frozen at the initial value $\theta_{\rm rh}$.
Around the time $t=t_{\rm osc,\pi}$ (black dashed vertical line, and as discussed above, the axion begins to oscillate around $\theta = \pi$, due to important temperature effects in its potential.
The QCD phase transition occurs at $t_c$ (gray dashed vertical line), at which point the temperature dependence of the axion potential quickly vanishes.
In this third epoch the axion energy density becomes dominated by its kinetic contribution, and the axion field rolls away from $\pi$.
At this point even very similar initial conditions will result drastically different evolutions for $\theta$, depending on the field velocity $\dot\theta_c$.
Finally, at $t_{\rm osc,0,}$ (black solid vertical line), the axion begins its second stage of oscillations, this time around $\theta = 2 \pi$ (or equivalently, $\theta = 0$), with a misalignment angle effectively given by $\theta_{\rm osc,0}$.
As discussed above, the preceding rolling era effectively ``shuffles'' the value of the axion angle, which can result in vastly different $\theta_{\rm osc,0}$ values even for very similar initial conditions.
In the bottom panel we show the evolution of the energy density of the axion, normalized by $t_c^2/f_a^2$.
The thick lines depict the total energy density, while the thin lines show only the kinetic term.
The blue, red, and green colors correspond to the same initial conditions as in the top panel.
In this panel we can easily see how, despite starting their evolution with very similar values, the axion energy density for each initial condition drastically changes soon after the QCD phase transition, due to the shuffling of the misalignment angle.
During the rolling phase, between the dashed gray and solid black vertical lines, the kinetic energy accounts for most of the axion energy density.

\paragraph{Rigged Misalignment:}
Let us now consider the opposite case and study the evolution of axions heavier than $H_c$, \ie~$m_a = \sqrt{\ephi}\mqcd\gtrsim H_c \approx 10^{-11}~\eV$. 
In this case, the axion begins to oscillation around the $\theta=0$ minimum immediately after the QCD phase transition, with initial amplitude
$\theta_c\sim\pi$ and a small velocity $\vert \dot\theta_c \vert \sim \mapi \vert \pi - \theta_c \vert$. 
Thus, unlike in the previous case, the field undergoes both oscillatory phases one right after the other, without an intervening rolling phase. 
In the absence of the rolling phase, the initial amplitude for the second phase of the oscillation is rigged to be $\sim \pi$ irrespective of the initial field value.  This behavior is why we called this {\it rigged misalignment}.   

As a result of rigging, the axion abundance in the late-time Universe can simply be obtained as 
\bea
\label{eq:mqcd_ma_gtr_hc}
\rho_a(t\gg t_c)\simeq \frac{\pi^2}{2}\ephi \mqcd^2 f_a^2 (t_c/t)^{3/2}\,,
\eea
which matches the DM energy density for 
\bea
\ephi\simeq 7.8\times 10^{-8} (r/1)\,. 
\eea
Notice that for a given $\ephi$, the estimate in Eq.~\eqref{eq:mqcd_ma_gtr_hc} is independent of $f_a$ (since $\mqcd f_a = \text{const.}$), as long as $\mqcd\gtrsim \mqcd\sqrt{\ephi}\gtrsim H_c$ \ie, for $f_a\lesssim 2\times 10^{14} \GeV (\ephi/7.8\times 10^{-8})^{1/2}$.

\begin{figure}[t!]
    \centering
    \begin{subfigure}{0.45\textwidth}
        \centering
        \includegraphics[width=\linewidth]{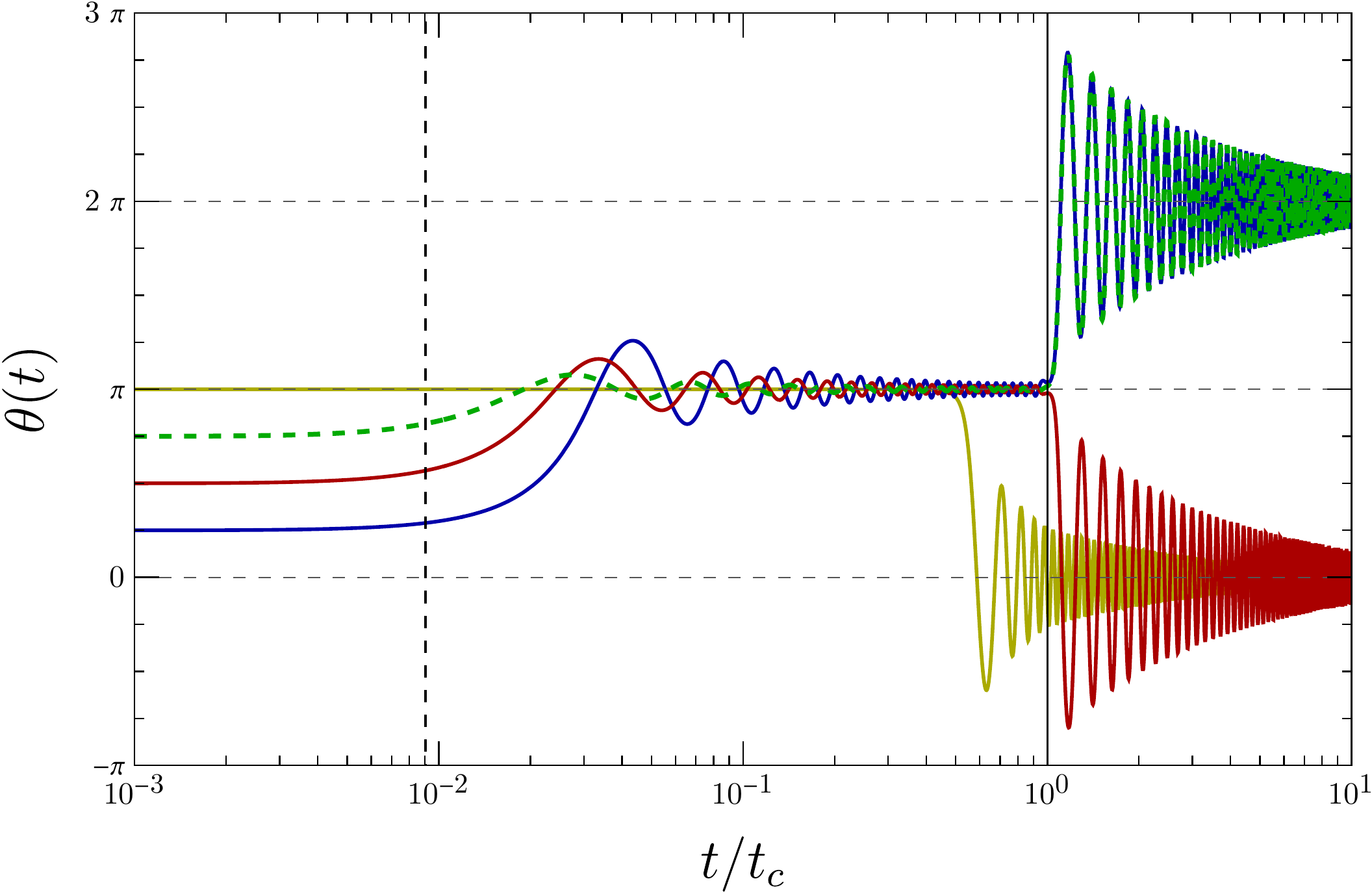}
    \end{subfigure}
    \begin{subfigure}{0.45\textwidth}
        \centering
        \includegraphics[width=\linewidth]{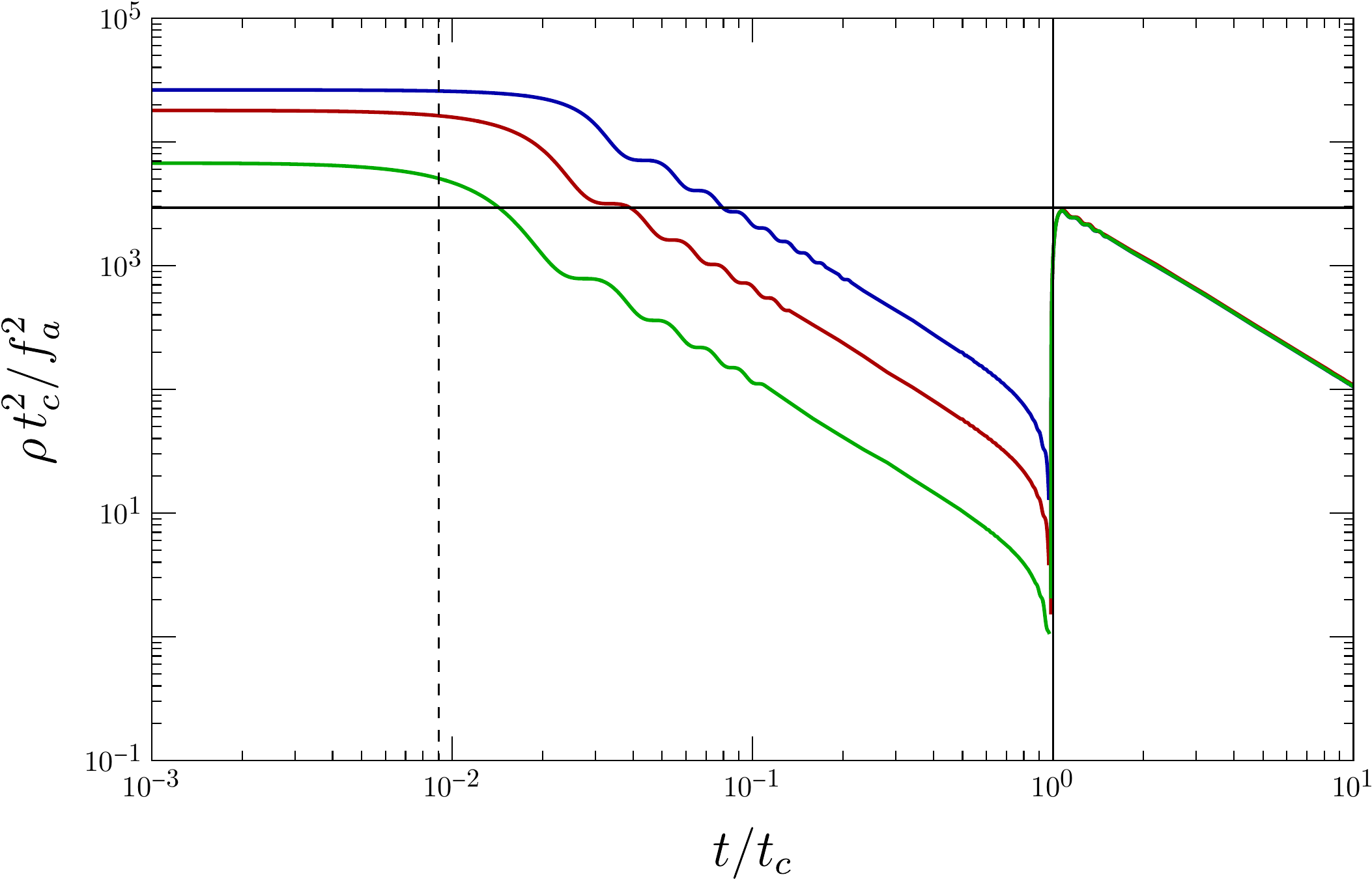}\\
    \end{subfigure}
    \caption{{\it Rigged misalignment}.
    The evolution of the axion angle $\theta$ ({\bf top}) and its normalized energy density $\rho_a \, t_c^2/f_a^2$ ({\bf bottom}) as a function of time $t$, for $f_a\simeq 2\times 10^{15}\GeV$ and $\ephi=0.1$.
    {\it Top:} The evolution of $\theta(t)$ for three very different initial conditions: $\theta_{\rm rh}= \pi/4$ (solid blue), $\theta_{\rm rh} = \pi/2$ (solid red), and $\theta_{\rm rh} = 3\pi/4$ (dashed green).
    The vertical and horizontal lines have the same meaning as those in Fig.~\ref{fig:case_1a}.
    Note that, due to the periodicity of the axion potential, the vertical axis in the top panel must be taken to be modulo $2\pi$.
    Therefore, the blue and green curves' oscillations around $\theta = 2\pi$ are the same as those of the red curve around $\theta = 0$.
    {\it Bottom:} The evolution of the energy density of the axion shifted by a constant such that when $t<t_c$, $\theta = \pi$ has zero energy and when $t>t_c$, $\theta = 0$ has zero energy. 
    The colors have the same meaning as in the top panel, and so do the vertical lines.  The black horizontal line is $\rho(\theta=\pi) - \rho(\theta=0)$.
    The fact that, regardless of initial conditions, the axion total energy density at $t=t_c$ jumps to black line demonstrates that it is restarting the misalignment mechanism with an angle rigged to $\theta = \pi$.
    }
    \label{fig:case_1b}
\end{figure}

In Fig.~\ref{fig:case_1b} we depict the evolution of the axion field ($\theta$, top panel) and its normalized energy density ($\rho_a t_c^2 / f_a^2$, bottom panel) in the rigged misalignment regime, with $f_a=2\times 10^{15}\GeV$ and $\ephi=0.1$.
The black dashed and solid vertical lines have the same meaning as those in Fig.~\ref{fig:case_1a}.
In the top panel, the three lines correspond to $\theta(t)$ with very different initial conditions: $\theta_{\rm rh} = \pi/4$ (solid blue), $\theta_{\rm rh} = \pi/2$ (solid red), and $\theta_{\rm rh} = 3\pi/4$ (dashed green).
As in Fig.~\ref{fig:case_1a}, the yellow line plots the canonical QCD axion, with same $f_a$ and $\theta_{\rm rh} = \pi-10^{-3}$. 
The two oscillatory eras described above, the first around $\theta = \pi$ and beginning at $t_{\rm osc,\pi}$, and the second around $\theta = 0$ (or equivalently $2\pi$) and beginning at $t_{\rm osc,0} = t_c$, can be easily seen in this figure.
Unlike in the shuffled misalignment case, however, the rolling phase is not present, and the two oscillatory phases take place in quick succession after the QCD phase transition.
Because of this the effective misalignment angle for the final oscillatory stage is $\theta_{\rm osc,0} = \theta_c \approx \pi$.
An initial misalignment angle so close to $\pi$ typically delays the onset of oscillations due to the flatness of the potential, as can be seen in the yellow line (see also refs.~\cite{Arvanitaki:2019rax,Chatrchyan:2023cmz}).
However, since in our model there is a non-zero field velocity at $t_c$, there is no such delay, and the axion field begins its second phase of oscillations immediately after the phase transition.

Since the effective misalignment angle of the last phase has been ``rigged'' to be $\approx \pi$ regardless of the initial conditions $\theta_{\rm rh}$, the subsequent evolution of the axion field and, consequently, of its energy density, is the same.
This can be seen by the overlap between the solid blue and dashed green lines.
The same is true of the red line which, despite appearing to be very different from the green and blue ones, is actually almost identical to them due to the mod $2\pi$ nature of the $\theta$ angle.

In the bottom panel we once again show the evolution of the energy density of the axion, normalized by $t_c^2/f_a^2$. 
The blue, red, and green colors correspond to the same initial conditions as in the top panel while the black horizontal line shows $\rho(\theta=\pi) - \rho(\theta=0)$.
In this panel we can easily see how, despite starting their evolution with very different values, the axion energy density for each initial condition all converge to the same evolution soon after the QCD phase transition, due to the rigged nature of the misalignment angle.
This can be explicitly seen when $t = t_c$, where the axion begins oscillating around $\theta = 0$ with a rigged effective initial misalignment angle of $\pi$. 

\paragraph{Folded Misalignment:}

So far, we have seen that for $\Trh \gtrsim T_c$ and $f_a \lesssim \fac$ the axion undergoes two oscillatory eras, whether in the shuffled or rigged regimes.
However it is possible for the axion to ``fold'', or not partake in the dynamics of this first stage for sufficiently large $f_a$ ($f_a \gtrsim \fac \approx 5\times 10^{16}\GeV$) and/or a small reheating temperature ($\Trh \lesssim T_c\sim 150\MeV$).
The axion field will always eventually oscillate around $\theta = 0$ with an initial amplitude of $\theta_{\rm osc,0}\simeq\theta_{\rm rh}$ as long as $m_a$ is larger than the Hubble expansion rate today. 
The oscillation time is obtained by solving $3H(T_{\rm osc,0})={\rm min}[3\Hrh,m_a]$ like the case of standard ALP dark matter.  Of course when $m_a > 3\Hrh$, $\theta_{\rm rh}$ will depend heavily on the dynamics preceding reheating. 
This case is similar to that of the standard misalignment for ALPs, and as such we defer a detail discussion of it to App.~\ref{app:other_cases}.

Let us conclude this section with a quick discussion about how requiring our axion to account for the totality of dark matter influences the parameter space.
Perhaps the most conservative demand we can make of an axion dark matter scenario is that it begins to oscillate no later than when the Universe had a temperature of $T_{\rm osc,0} \sim \keV$, which corresponds to the time when the smallest modes probed by current CMB observations entered the horizon~\cite{Hui:2016ltb}.
A quick calculation shows that such a constraint translates into a bound of $\ephi\gtrsim 4\times 10^{-21}$ for $f_a=\Mpl$, not very far from the independent bound on $\ephi$ outlined in Subsec.~\ref{subsec:thermal}.

\section{Discussion and Conclusion}\label{sec:conclusion}

\begin{figure}[t!]
    \centering
    \includegraphics[width=\linewidth]{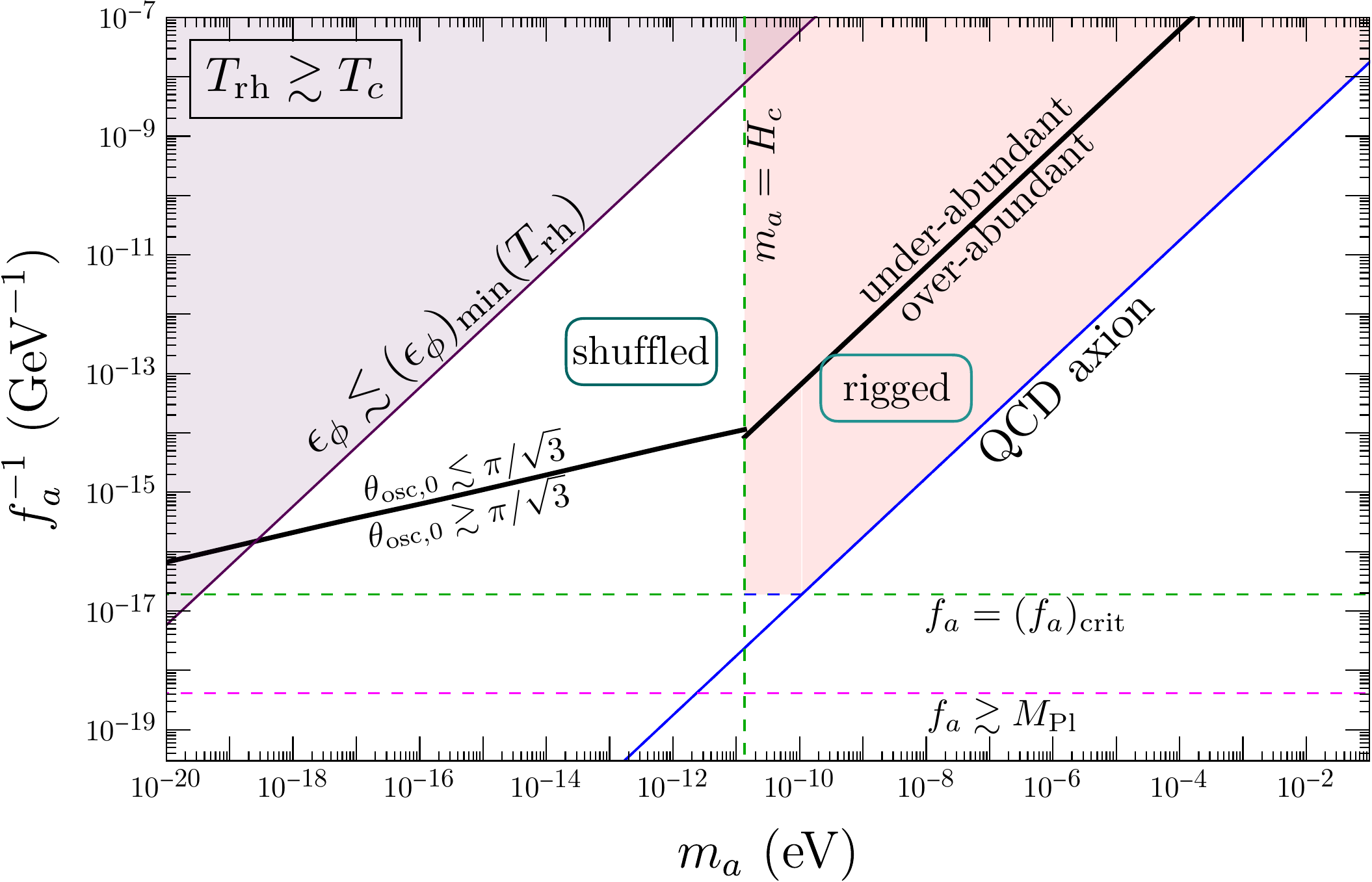}
    \caption{The $m_a$--$f_a$ parameter space of our model, for the $\Trh \gtrsim T_c$ case.
    Here we have taken $\Trh = 1~\GeV$.
    The dashed green and pink horizontal lines correspond to $f_a = \fac$ and $f_a = M_{\rm Pl}$, respectively.
    The blue line corresponds to the canonical QCD axion.
    The brown region is ruled out by various astrophysical and cosmological bounds.
    The dashed green vertical line corresponds to $m_a = H_c$, which represents the boundary between the shuffled and rigged regimes of the two-stage misalignment mechanism of our model.
    Since $\ephi < 1$, both regimes lie to the left of the blue line.
    In the purple region $\ephi \lesssim (\ephi)_{\rm min}$ as given by Eq.~\eqref{eq:ephi_Trh}.
    For this plot, $(\ephi)_{\rm min} \approx 9 \times 10^{-20}$, for $m_\phi = \Lnp = 3\,\TeV$, and $\Trh = 1\,\GeV$.
    The black line corresponds to those values of $f_a$ and $m_a$ for which the axion can account for the entirety of the dark matter.
    In the shuffled regime, one can still get the right dark matter abundance above or below this line by tuning $\theta_{\rm osc,0}$ in the ways indicated in the plots.
    In the rigged regime, all points above (below) the black line, shown by the light red shading, correspond to an under-abundance (overabundance) of axion dark matter.
    We have omitted the experimental bounds on this parameter space to avoid overcrowding the plot; they can be seen in Fig.~\ref{fig:DM_small_Trh}.
    }
    \label{fig:DM_large_Trh}
\end{figure}

\begin{figure}[t]
    \centering
    \includegraphics[width=\linewidth]{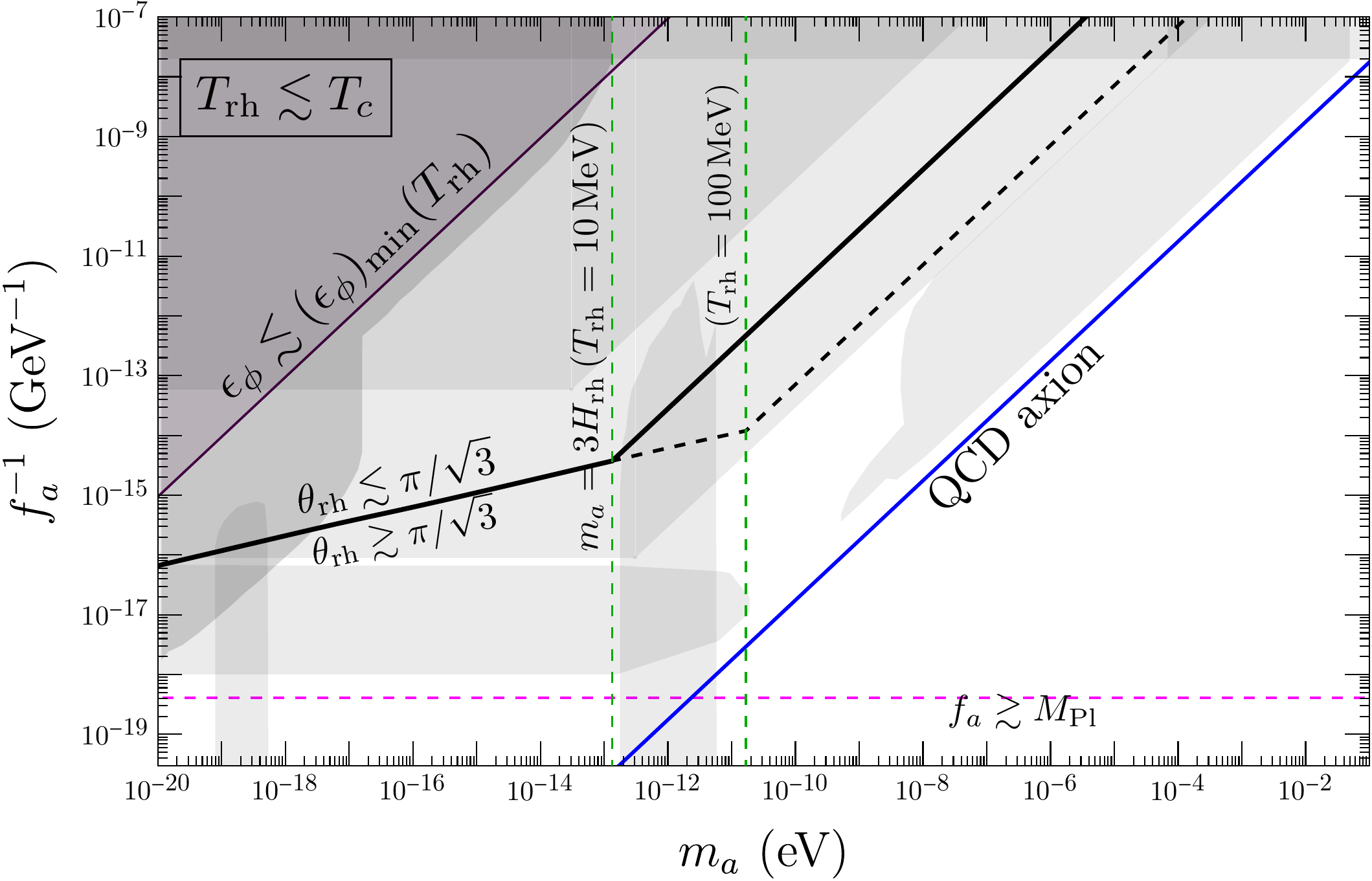}
    \caption{The parameter space of our model when $\Trh \lesssim T_c$. 
    In this case, there is no first stage of oscillations around $\theta = \pi$, and thus there is neither a shuffled nor a rigged regime.
    The dashed pink and blue lines and brown region have the same meaning as those in Fig.~\ref{fig:DM_large_Trh}.
    Since $\ephi < 1$, our model lies to the left of the blue line.
    The dashed green vertical lines correspond to $m_a = 3 H_{\rm rh}$, for $\Trh = 10\,\MeV$ and $\Trh = 100\,\MeV$.
    These lines represent the boundary between when the axion oscillates some time afterward reheating (left of the line) or the more model dependent case where it starts oscillating before or at reheating (right of the line).
    In the purple region $\ephi \lesssim (\ephi)_{\rm min}$ as given by Eq.~\eqref{eq:ephi_Trh}.
    For this plot, $(\ephi)_{\rm min} \approx 3 \times 10^{-24}$, for $m_\phi = \Lnp = 3\,\TeV$, and $\Trh = 10\,\MeV$.
    The black line corresponds to those values of $f_a$ and $m_a$ for which the axion can account for the entirety of the dark matter.
    One can still get the right dark matter abundance above or below this line depending on the value of $\theta_{\rm rh}$. 
    The gray shaded regions represent combined exclusions from various model dependent astrophysical, cosmological, and laboratory experimental results~\cite{Hook:2017psm,Zhang:2021mks,Schulthess:2022pbp,Blum:2014vsa,Gue:2025nxq,Zhang:2022ewz,JEDI:2022hxa,Roussy:2020ily,Abel:2017rtm,Zhang:2023lem,Alda:2024xxa,Banerjee:2023bjc,Gomez-Banon:2024oux,Kumamoto:2024wjd,Balkin:2022qer,Hook:2017psm,Lucente:2022vuo,Springmann:2024ret,Caloni:2022uya,Baryakhtar:2020gao,Baryakhtar:2020gao,Stott:2020gjj,Unal:2020jiy,Hoof:2024quk,Witte:2024drg,Caputo:2024oqc,Iwamoto:1984ir,Buschmann:2021juv,Springmann:2024mjp, Springmann:2024ret}. 
    }
\label{fig:DM_small_Trh}
\end{figure}

In Figs.~\ref{fig:DM_large_Trh} and~\ref{fig:DM_small_Trh} we show our model's $m_a$--$f_a$ parameter space, for $\Trh \gtrsim T_c$ and $\Trh \lesssim T_c$, respectively.
In Fig.~\ref{fig:DM_large_Trh} we took $\Trh=1 \GeV$; its precise value has no impact on the late time abundance of the axion as long as it is above $T_c$.
On the other hand, we allow $\Trh$ to vary between $10\,\MeV$ and $100\,\MeV$ in Fig.~\ref{fig:DM_small_Trh}.
In both figures, the dashed green and pink horizontal lines correspond to $f_a = \fac$ and $f_a = M_{\rm Pl}$, respectively.
The canonical QCD axion is shown as a blue line \cite{Weinberg:1977ma,Wilczek:1977pj,Kim:1979if,Shifman:1979if,Zhitnitsky:1980tq,Dine:1981rt}.
Since $\ephi \lesssim 1$, our model lives to the left of this blue line.
The gray shaded regions represent combined exclusions from various model dependent astrophysical, cosmological, and laboratory experimental results~\cite{Hook:2017psm,Zhang:2021mks,Schulthess:2022pbp,Blum:2014vsa,Gue:2025nxq,Zhang:2022ewz,JEDI:2022hxa,Roussy:2020ily,Abel:2017rtm,Zhang:2023lem,Alda:2024xxa,Banerjee:2023bjc,Gomez-Banon:2024oux,Kumamoto:2024wjd,Balkin:2022qer,Hook:2017psm,Lucente:2022vuo,Springmann:2024ret,Caloni:2022uya,Baryakhtar:2020gao,Baryakhtar:2020gao,Stott:2020gjj,Unal:2020jiy,Hoof:2024quk,Witte:2024drg,Caputo:2024oqc,Iwamoto:1984ir,Buschmann:2021juv,Springmann:2024mjp, Springmann:2024ret}. 
To avoid clutter, these constraints are only shown in Fig.~\ref{fig:DM_small_Trh} but not shown in Fig.~\ref{fig:DM_large_Trh}, however the constraints apply equally well in both scenarios.
In purple, we shade the region for which $\ephi$ is so small that the desired reheat temperature cannot be obtained ($\ephi \lesssim (\ephi)_{\rm min}$ as given by Eq.~\eqref{eq:ephi_Trh}).
Taking $m_\phi = \Lnp = 3\,\TeV$, we get $(\ephi)_{\rm min} \approx 9 \times 10^{-20}$ for $\Trh = 1\,\GeV$ in Fig.~\ref{fig:DM_large_Trh}, and $(\ephi)_{\rm min} \approx 3 \times 10^{-24}$, for $\Trh = 10\,\MeV$ in Fig.~\ref{fig:DM_small_Trh}.
In Fig.~\ref{fig:DM_large_Trh}, the dashed green vertical line corresponds to $m_a = H_c \approx 10^{-11}~\eV$, which roughly separates shuffled and rigged misalignment regimes.

In the rigged regime, the black line indicates when the axion can account for the entirety of the dark matter.  Eq.~\eqref{eq:mqcd_ma_gtr_hc} tells us that, since $\theta_{\rm osc,0} \approx \pi$, the axion dark matter will be underabundant or overabundant above or below this black line.
In the shuffled regime, the black line indicates when the axion can account for the entirety of the dark matter for a typical misalignment angle of $\theta_{\rm osc,0} = \pi/\sqrt{3}$.
Eq.~\eqref{eq:fa_adm_shuffle} indicates that the correct DM abundance can still be obtained above or below the black line, by choosing initial conditions such that $\theta_{\rm osc,0}$ lies closer to $\theta = 0$ or to $\theta = \pi$, respectively.

In Fig.~\ref{fig:DM_small_Trh} as $\Trh \lesssim T_c$, the axion skips the first stage of oscillations around $\theta = \pi$, and its evolution is like that of a light ALP. 
The dashed green lines correspond to the transition between the axion oscillating some time after reheating (to the left of the vertical lines) to oscillating sometime before reheating takes place (to the right of the vertical lines) for $\Trh = 10\,\MeV$ and $100\,\MeV$.  To avoid discussing the details of reheating, we present our plots in terms of $\theta_{\rm rh}$ with the understanding that its value will depend on the unspecified dynamics occurring before reheating.
As in the previous figure, the black line corresponds to points in the parameter space for which the axion dark matter has the observed abundance for a typical misalignment angle of $\theta_{\rm rh} \approx \pi/\sqrt{3}$; tuning this angle closer to $0$ or $\pi$ allows the axion to be the dark matter above or below this line.


In this work we demonstrated that by introducing a soft $\mZ_N$‐breaking coupling, one can simultaneously populate the light QCD‐axion parameter space up to masses $\mathcal{O}(10^{-12})$ times lighter than their canonical QCD values, while resolving the usual $1/N$ tuning and reheating ambiguities of the original $\mZ_N$ QCD axion model.
We extended the $\mZ_N$ QCD axion framework by introducing a $\mZ_N$‐symmetry breaking coupling, $g$, between the Higgs and the reheaton field ($\phi$), through which the reheaton decays exclusively into our Standard Model sector.
This coupling also induces a slight shift in the Higgs VEV in our sector compared to the others.
As a result, the sector that gets reheated has a slightly larger axion potential, leading to an imperfect $\mZ_N$ cancellation that selects the $\theta=0$ minimum.

Due to the shifted Higgs VEV, the axion mass becomes $m_a^2 \simeq (\ephi+\eps_N) \, \mqcd^2$, where $\eps_N$ is the axion mass in the $\mZ_N$ limit.
The presence of $\ephi \sim g^2/m_\phi^2$ decouples the axion mass from its decay constant $f_a$, allowing the model to accommodate a lighter axion for any $\eps_N\lesssim\ephi\ll 1$.
Thus, unlike the original $\mZ_N$ construction--- which predicts a discrete set of axion masses for odd values of $N$---our setup continuously populates the axion parameter space for arbitrary $N$. 

Cosmologically, this setup gives rise to three distinct misalignment regimes depending on the reheating temperature and the axion decay constant.
For high reheating temperatures ($\Trh \gtrsim$ GeV) and small $f_a\lesssim 5\times 10^{16}\GeV$, early oscillations around $\theta=\pi$
pin the axion near $\pi$ until the QCD phase transition. 
For heavier axions with $m_a \gtrsim H_c \approx 10^{-11}\eV$, the field begins oscillating around the $\theta=0$ minimum immediately after the QCD phase transition.
This dynamically fixes the late‐time misalignment angle---and therefore the relic abundance---along a single trajectory in the $(m_a,f_a)$ plane, regardless of the initial condition at the time of reheating.
We refer to this regime as ``rigged" misalignment.
In contrast, for $m_a \lesssim H_c \approx 10^{-11}\eV$, the axion undergoes a rolling phase after initially oscillating around $\pi$. 
This rolling randomizes the effective misalignment angle, allowing for the usual ALP‐like tuning to reproduce the observed dark matter (DM) energy density. 
We refer to this as the ``shuffled'' regime. 
For $f_a\gg 5\times 10^{16}\GeV$ and/or low reheating temperatures, the axion skips the initial oscillations around $\theta=\pi$, resulting in the standard ALP dark matter scenario.
We refer to this as ``folded" misalignment. 

In this work, we do not construct a UV complete model that simultaneously generates the $\mathbb{Z}_N$ symmetric coupling at the scale $f_a$, and naturally explains a small symmetry-breaking Higgs-reheaton coupling at a lower scale. 
While such symmetry-breaking features can be readily embedded within a KSVZ-type UV completion of the $\mathbb{Z}_N$ axion, we defer a detailed exploration of this possibility to future work. 
Also currently in our model, we focus is on providing analytical estimates for various dark matter scenarios without performing full numerical simulations. In particular, in the ``rigged'' misalignment regimes—where the axion begins with an initial misalignment angle of $\mathcal{O}(\pi)$-a detailed understanding of dark matter perturbation growth and structure formation may require incorporating non-linear dynamics and topological effects, which lie beyond the scope of this study. 
Finally, a thorough analysis of the full temperature-dependent axion potential during a kination phase could unveil novel cosmological signatures. We leave this investigation to future work as well.

\acknowledgements{}

The authors would like to thank Sebastian A.R. Ellis for providing us with tabulated data of the existing constraints. 
The authors would also like to thank Michael Geller, and Konstantin Springmann for useful discussions and reading the draft. 
The authors are supported by the National Science Foundation under grant number PHY2210361 and the Maryland Center for Fundamental Physics.
AB would also like to acknowledge the Tatte Bakery at the City Center, where a good chunk of the work was done.

\appendix

\renewcommand{\theequation}{\thesection.\arabic{equation}}
\setcounter{equation}{0}

\section{Quality Problem, $Z_N$ and light QCD axion}
We devote this appendix to the status of the quality problem in our model.
In the framework of EFT, one generically expects that the global Peccei-Quinn (PQ) symmetry, which gives rise to the axion, is broken by quantum gravity effects~\cite{Kamionkowski:1992mf,Barr:1992qq,Ghigna:1992iv}, although with the expectation that such effects decouple in the limit $\Mpl\to \infty$.

Let us begin by discussing the quality problem of the canonical QCD axion. 
In this case one can, for instance, write operators of the form
\bea
V_{\Delta} &=& \left(\frac{c_p \Phi^{p+4}}{\Mpl^{p}} +{\rm h.c.}\right) + (\text{similar order terms}) \nn\\
&\rightarrow &\left(\frac{f_a}{\Mpl}\right)^{p} f^4_a \cos(p\theta+ \beta)\,,
\label{eq:QCd_axion_quality}
\eea
which explicitly breaks the PQ symmetry and generates a potential for the QCD axion.
Here $\beta= {\rm Arg}(c_p)$ is a generic phase, which is set to  $\mathcal{O}(1)$. 
In general, these Planck suppressed (``quality") operators shift the minima of the axion potential, which leads to a constraint on $p$:
\bea
    p\gtrsim \frac{125 + 4 \log (f_{a}/10^{10}\GeV)}{19- \log (f_a/10^{10}\GeV)}\,,
\eea
in order for the leading operator in Eq.~\eqref{eq:QCd_axion_quality} not to spoil the axion solution to the strong CP problem and generate a neutron EDM.

\begin{figure}[t]
    \centering
    \includegraphics[width=\linewidth]{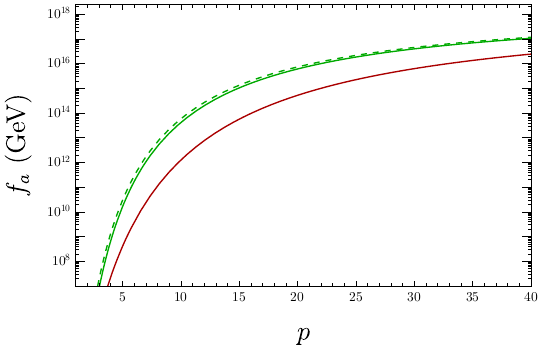}
    \caption{Comparison of the dimension of the quality operators for the canonical QCD axion (red) and our model of the lighter QCD axion (green). 
    We see that for a given $f_a$, despite being lighter than the canonical QCD axion, the axion in our model requires less protection from the Planck suppressed operators than the canonical QCD axion case. 
    For the solid lines, we take $\Lnp=m_\phi=3\TeV$,  whereas for the dashed line we take $\Lnp=100\TeV$ and $m_\phi=10\TeV$. 
    }
    \label{fig:quality_comparison}
\end{figure}

In our mode, where the QCD axion is lighter than usual, one might naively expect that the quality of the axion is worsened, since the discrete $\mZ_{N}$ symmetry is broken softly by $g$.
However, since the underlying  $\mZ_{N}$ symmetry is restored as $g\to 0$, it can be gauged, which means that the leading-order operators spoiling the axion quality should depend on $g$.
More importantly, since $(g,\phi)\to (-g,-\phi)$ is still a symmetry of the system, these leading-order operators should be $\propto g^2$.
Therefore, in our model, the leading PQ symmetry breaking operator takes the form of
\bea
V_{\Delta} &=& g^2 \left(\frac{c_p \Phi^{p+2}}{\Mpl^{p}} +{\rm h.c.}\right) + (\text{similar order terms}) \nn\\
&\rightarrow &\left(\frac{f_a}{\Mpl}\right)^{p} g^2 f^2_a \cos(p\theta+ \beta)\,.
\label{eq:light_axion_quality}
\eea
For $\ephi \gtrsim \eps_N$, using Eq.~\eqref{eq:potential_T_0} and Eq.~\eqref{eq:ephi_def},  we find that these operators do not run afoul neutron EDM experiments for
\begin{align}
    p\gtrsim \frac{93 - 2 \log[(\Lnp/m_\phi)] + 2\log(f_a/10^{10}\GeV)}{19 -\log(f_a/10^{10}\GeV) }\,.
\end{align}
In Fig.~\ref{fig:quality_comparison} we show the quality of our light axion by green lines, and depict the quality of the canonical QCD axion in red. 
As can be seen from Eq.~\eqref{eq:light_axion_quality}, the value of $p$ for a given $f_a$ depends on $\ephi/g^2\propto \Lnp^2/m_\phi^2$ for $\Lnp\gtrsim v_{k \neq 0}$. 
The solid line shows the value of $p$ for a given $f_a$ for $m_\phi/\Lnp=1$, whereas the dashed line depicts for $m_\phi/\Lnp=0.1$. 
Compared to the case of a canonical QCD axion, the quality of our light axion is better, as long as $\Lnp f_a/(v_{k\neq0} m_\phi)\gtrsim 1$. 
Since $m_\phi\lesssim \Lnp$, we obtain a better quality axion for any $f_a\gtrsim v_0$. 
Finally, we note that the effects of $\mZ_N$-symmetric, Planck suppressed operators are small, since they are suppressed by powers of $(f_a/\Mpl)^N$ (for a specific UV completion), which is negligible in the large-$N$ limit in which we are interested (see {\it e.g.}~\cite{Banerjee:2022wzk} for a discussion).

\section{Folded Misalignment}
\label{app:other_cases}

In this section, we discuss the evolution of the axion in the ``folded" misalignment region of parameter space. 
As mentioned in the main text, folded misalignment occurs for large $f_a$ ($f_a\gtrsim \fac \approx 5\times 10^{16}\GeV$) and/or a small reheating temperature ($\Trh\lesssim T_c\sim 150\MeV$).
In these cases, the axion DM folds on any complicated new dynamics and is produced via the regular misalignment mechanism, similar to the case of ALPs.  
\\
\paragraph{Large $f_a \gtrsim 5\times 10^{16}\GeV$\,: } 

If the axion decay constant is large \ie~$f_a\gg \fac$, then regardless of whether or not the reheat temperature was larger than the QCD scale, the field does not undergo the first era of oscillations, since for such $f_a$ values $ H(t_c) \gg \mqcd$. 
The field remain frozen for a long time, and only starts to oscillate around $\theta = 0$ at $T_{\rm osc,0}$ when $3 H(T_{\rm osc,0}) = m_a=\mqcd \sqrt{\ephi}\,$. 
In this case, the high temperature potential plays no role, and $\theta_{\rm osc,0} \approx \theta_{\rm rh}$. 
The axion accounts for the present day dark matter energy density for 
\begin{align}
\label{eq:fa_ALP_DM}
f_a\simeq  10^{17}\GeV \left(\frac{\eps_\phi}{8\times 10^{-17}}\right)^{-\frac{1}{6}} \left(\frac{\theta_{\rm rh}}{0.1}\right)^{-\frac{4}{3}}\!\!\!\left(\frac{r}{1}\right)^{\frac{2}{3}}\,.
\end{align}
Thus a mild tuning of the initial condition is needed for the axion to account for $100\%$ of the dark matter.

We finish our discussion of this scenario by noting that, for  $f_a\sim \fac$ the first oscillatory phase is replaced by a rolling phase, in which the field slowly evolves towards $\theta = \pi$.
Here the field reaches $\theta = \pi$ only if $\mqcd/H(T_c) \sim \cO(5)$.
In general the axion will either roll straight past $\pi$, or undergo at most half an oscillation before rolling past it.
More importantly, the field is effectively displaced from its initial value, and the misalignment angle for the second phase of oscillation gets ``shuffled'', \ie, ends up being very different from its original value. 
We already consider this case while plotting the parameter space of our model in Fig.~\ref{fig:DM_large_Trh}. \\

\paragraph{$\Trh \lesssim T_c$:} 
For a reheating temperature smaller than $T_c$, the temperature corrections to the axion potential are essentially negligible. 
Thus, the axion potential is always minimized at $\theta=0$, and its mass is always given by its $T=0$ value. 
Unlike the case of $\Trh\gtrsim T_c$, the absence of the temperature effects ensure that the axion does not undergo its earlier oscillatory and/or rolling phases prior to oscillating around $\theta=0$.
Once again, this case is similar to that of the standard ALP dark matter; the field starts to oscillate at $t_{\rm osc,0}$ which is obtained by solving $H(T_{\rm osc,0})={\rm min}[\Hrh,m_a/3]$ with an initial amplitude of $\theta_{\rm osc,0}\simeq\theta_{\rm rh}$. 
When $m_a \gtrsim 3 \Hrh$, $\theta_{\rm rh}$ becomes extremely sensitive to the unspecified cosmological history before reheating.
When $H(T_{\rm osc,0})=\Hrh$ \ie for large $m_a$, the axion abundance becomes independent of $f_a$, and its parametric dependence is given as $\rho_a\propto \eps_\phi/\Trh^3$.
We find that the axion relic constitutes 100\% of the present day dark matter energy density for
\bea
\!\!\!\!\!\!\!\!\!
\epsilon_\phi=2\times 10^{-8}  \left(\frac{\Hrh}{H(0.1\GeV)}\right)^{3/2}\left(\frac{1}{\theta_{\rm rh}}\frac{r}{1}\right)^{2},
\eea
for $f_a\lesssim 4.8\times 10^{13}\GeV \sqrt{\eps_\phi/(2\times 10^{-8})}/(\Hrh/0.1\GeV)$.
Since $\Trh\gtrsim 10\MeV\gtrsim T_{\rm BBN}\sim \MeV$, this occurs for 
\bea
f_a\lesssim 6\times 10^{15}\GeV\sqrt{\frac{\eps_\phi}{2\times 10^{-8}}}
\,.
\eea
On the other hand, if $\Hrh\gtrsim m_a/3$, the axion starts to oscillate when $3H(T_{\rm osc,0}) = m_a$, and its abundance follows the scaling given in Eq.~\eqref{eq:fa_ALP_DM}.

\bibliography{ref.bib}
\end{document}